\documentclass[10pt,aps,pra,twocolumn,superscriptaddress,floatfix,nofootinbib]{revtex4-2}

\usepackage{amsmath, amsfonts, amssymb}
\usepackage{bm,bbm, braket, accents}
\DeclareMathOperator{\sech}{sech}
\usepackage[usenames,dvipsnames]{color}
\usepackage{lipsum} 
\usepackage{tikz}
\usetikzlibrary{quantikz}
\usepackage{pgfplots}
\pgfplotsset{compat=newest}

\usepackage{dcolumn}% Align table columns on decimal point
\usepackage{bm}% bold math
\usepackage{soul}
\makeatletter
\def\@bibdataout@aps{%
 \immediate\write\@bibdataout{%
  @CONTROL{%
   apsrev41Control,author="08",editor="1",pages="0",title="0",year="1",eprint="1"%
  }%
 }%
 \if@filesw
  \immediate\write\@auxout{\string\citation{apsrev41Control}}%
 \fi
}%
\makeatother % Phew.

\newcommand{\<}{\langle}
\renewcommand{\>}{\rangle}

\newcommand\vac{\mathrm{vac}}
\newcommand{\lo}{\text{LO}}
\newcommand{\sig}{\text{S}}

\newcommand{\app}{Appx. }

\usepackage[breaklinks=true]{hyperref}
\hypersetup{
  colorlinks   = true, %Colour links instead of boxes
  urlcolor     = blue, %Colour for external hyperlinks
  linkcolor    = blue, %Colour of internal links
  citecolor    = red %Colour of citations
}

\newcommand\red{\color{red}}

\usepackage{cleveref} % must be loaded after hyperref
\crefformat{equation}{Eq.~(#2#1#3)} % These change 'equation' to Eq., more PRL or PRA-style
\crefformat{section}{Sec.~#2#1#3} % These change 'equation' to Eq., morePRL or PRA-style
\Crefformat{equation}{Equation~(#2#1#3)}
\crefformat{figure}{Fig.~#2#1#3}
\crefrangeformat{equation}{Eqs.~#3(#1)#4--#5(#2)#6}
\Crefformat{section}{Section~#2#1#3}
\crefformat{appendix}{App.~#2#1#3}
\usepackage{graphicx}   % need for figures

\begin{document}

\title{Quantum theory of temporally mismatched homodyne measurements with applications to optical frequency comb metrology}

\author{Noah Lordi}
\altaffiliation{These two authors contributed equally and can be reached at noah.lordi@colorado.edu and eugene.tsao@colorado.edu.}
\affiliation{Department of Physics, University of Colorado Boulder, 2000 Colorado Ave, Boulder, CO 80309
}

\author{Eugene J. Tsao}%
\altaffiliation{These two authors contributed equally and can be reached at noah.lordi@colorado.edu and eugene.tsao@colorado.edu.}

\affiliation{Electrical, Computer, and Energy Engineering, University of Colorado Boulder, 1111 Engineering Dr, Boulder, CO 80309
}
\affiliation{Time and Frequency Division, National Institute of Standards and Technology, 325 Broadway, Boulder, CO 80305
}%

\author{\\ Alexander J. Lind}%

\affiliation{Electrical, Computer, and Energy Engineering, University of Colorado Boulder, 1111 Engineering Dr, Boulder, CO 80309
}
\affiliation{Department of Physics, University of Colorado Boulder, 2000 Colorado Ave, Boulder, CO 80309
}
\affiliation{Time and Frequency Division, National Institute of Standards and Technology, 325 Broadway, Boulder, CO 80305
}%

\author{Scott A. Diddams}%

\affiliation{Electrical, Computer, and Energy Engineering, University of Colorado Boulder, 1111 Engineering Dr, Boulder, CO 80309
}%
\affiliation{Department of Physics, University of Colorado Boulder, 2000 Colorado Ave, Boulder, CO 80309
}
\affiliation{Time and Frequency Division, National Institute of Standards and Technology, 325 Broadway, Boulder, CO 80305
}%

\author{Joshua Combes}%

\affiliation{Electrical, Computer, and Energy Engineering, University of Colorado Boulder, 1111 Engineering Dr, Boulder, CO 80309
}%

\date{\today}% It is always \today, today,

\begin{abstract}
The fields of precision timekeeping and spectroscopy increasingly rely on optical frequency comb interferometry. However, comb-based measurements are not described by existing quantum theory because they exhibit both large mode mismatch and finite strength local oscillators.
To establish this quantum theory, we derive measurement operators for homodyne detection with arbitrary mode overlap. 
These operators are a combination of quadrature and intensity-like measurements, 
which inform a filter that maximizes the quadrature measurement signal-to-noise ratio. 
Furthermore, these operators establish a foundation to extend frequency-comb interferometry to a wide range of scenarios, including metrology with nonclassical states of light. 
\end{abstract}

\maketitle

\section{Introduction}
Homodyne measurements~\cite{HuntingtonLovovsky2012} are foundational to quantum optics and precision metrology, enabling the manipulation~\cite{Gerrits2011,Fuwa2015} and characterization~\cite{Polycarpou2012,Qin2015,GrandiParis2017} of quantum states. In a single mode, homodyne is understood as a measurement of a quadrature of the electromagnetic field~\cite{Walker87,Braunstein90,BanaszekWodkiewicz1997,Tyc2004}. Many multi-mode formulations of homodyne assume the signal and local oscillator (LO) share the same temporal mode~\cite{Helstrom1967,Shapiro85,Collett87,Blow1990,Bennink2002,Nakamura2021} resulting in qualitatively similar quantum descriptions and limits as the single mode case. Temporal mode mismatch between the signal and LO has been understood as effective loss \cite{Grosshans2001,Polycarpou2012,Qin2015,ou2017quantum,ChenWangYu2023}. However, these works do not consider the effects of finite strength LOs, large signal strength, and large mode mismatch evident in many experiments~\cite{Deschenes2013,Deschenes2015,Walsh2023}. In fact, these experiments observe additional shot noise due to mode mismatch, which is unexplained by effective loss alone.

Over the past two decades, optical frequency combs \cite{Diddams2020} have emerged as a powerful tool for characterization and dissemination of the most precise clocks \cite{Beloy2021,Caldwell2022,Shen2022}, precision spectroscopy \cite{Coddington2016,Picque} and broad bandwidth frequency synthesis \cite{Yao2016,Xie2016,Nakamura2020b}. 
In these measurements, the frequency comb LOs have very high peak power, but relativity small average power; as a result the finite strength LO effects are important particularly when the signal has similar average power to the LO.  As comb-based measurements near putative shot-noise quantum limits \cite{REICHERT1999,Deschenes2013,Caldwell2022,Quinlan2013}, a quantum measurement description that addresses temporal mode mismatch and finite field strengths is crucial to determine the fundamental bounds on precision. A complete quantum theory also forms the foundation of frequency comb metrology with non-classical light~\cite{PinelJianTreps2012, KuesMorandotti2019, FabreMilman2020, MalteseDucci2020}. We expect this to be important for comb-based measurements aimed at surpassing the standard quantum limit \cite{YangYi2021,CaiRoslundTreps2021,Belsley2023,GuidryVuckovic2023,ShiChenZhuang2023}.

\begin{figure}[!t]
    \centering
    \includegraphics[width=\columnwidth]{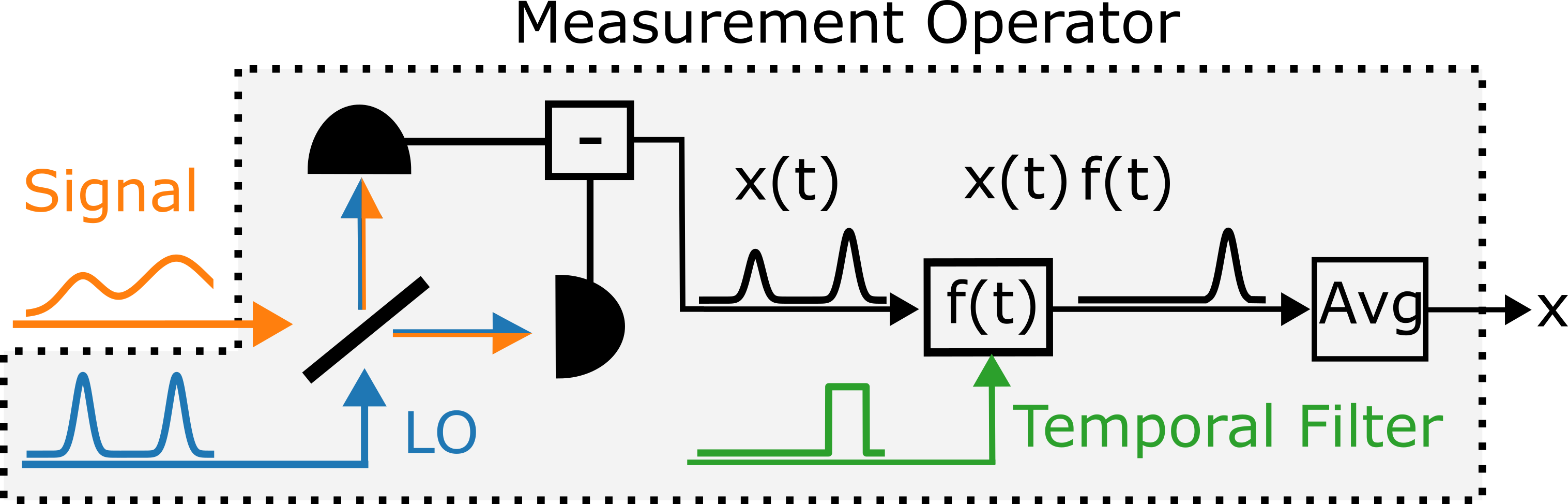}
    \caption{Temporally mismatched homodyne measurement with signal processing. The signal and local oscillator (LO) have different temporal modes, resulting in a mismatch. The detectors produce a photocurrent proportional to the intensity of the field and the instantaneous difference is denoted by $x(t)\propto I_1(t)-I_2(t)$. We compute the measurement operators for this setup and consider filtering the photocurrent. This filtering removes the additional shot noise due to mode mismatch~\cite{Deschenes2013}.  We describe the set of filters that do not affect the measured quadrature and achieve SNRs considerably larger than the unfiltered SNR.
    }\label{fig:apparatus}
\end{figure}

 In this Article, we address this need by providing a quantum description of temporally mismatched homodyne measurement, shown in \cref{fig:apparatus}. Specifically, we derive the measurement operators, i.e., the 
 positive operator valued measures (POVMs), which enable the calculation of measurement statistics for any signal state and coherent LO both with arbitrary time dependence.

This article is organized as follows. In \cref{sec:continous_modes} we use continuous mode quantum optics and the Gram-Schmidt procedure to decompose the mode of an incoming signal into the mode of the LO and an orthogonal mode.
This temporal mode decomposition is used to derive measurement operators for modal homodyne in \cref{sec:modal_homodyne}, which is our main result. The measurement operator consists of two parts: a quadrature measurement (corresponding to the LO mode) and an intensity-like measurement (corresponding to the orthogonal mode).  We then illustrate our formalism in \cref{sec:examples} with several examples. Specifically we use our analysis to develop a new quantum limit for comb-based measurement and provide quantum theoretic grounds for experimental results demonstrating better-than-shot noise-limited performance by Deschênes and Genest \cite{Deschenes2013} via temporal filtering. We also present measurement statistics for an example non-classical signal that--to our knowledge--cannot be described by existing analyses. Finally, we conclude in \cref{sec:conclusion} with a discussion of the implications of our results for heterodyne, which is the standard measurement with frequency combs.

%==================================
\section{Continuous modes and Gram-Schmidt}\label{sec:continous_modes}
%==================================
In a balanced homodyne measurement, the signal and local oscillator (LO) are combined on a beamsplitter and both output ports are detected. The resulting photo-currents are then subtracted, and the difference is recorded as the measurement result (see \cref{fig:apparatus}). Typically, the LO strength dominates the signal strength and the LO is temporally mode-matched to the signal. Here we do not assume that the signal and LO are mode-matched and allow for arbitrary mode overlap. For this reason we need to introduce the basics of continuous mode quantum optics~\cite{Blow1990}.

We begin by defining the mode creation operator $\hat{A}^\dagger(\xi)$ in some temporal mode $\xi(t)$, also known as the field envelope,
\begin{equation}
    \hat{A}^\dagger(\xi) = \int_0^T dt\,\xi(t)\hat{a}^\dagger(t)\, ,
\end{equation}
where $\hat{a}^\dagger(t)$ is the creation operator that creates a photon at time $t$. These mode operators carry the usual commutation relations $[\hat{A}(\xi),\hat{A}^\dagger(\xi)] = 1$ unlike the instantaneous creation operators which have units of ${\text{sec}^{-1/2}}$ as can be seen from  $[\hat{a}(t), \hat{a}^\dagger(t')]=\delta(t-t')$.

To describe time-dependent homodyne measurements we introduce independent and arbitrary complex temporal modes for the signal and LO, denoted by the mode functions $\xi_\sig(t)$ and $\xi_\lo(t)$. These modes are normalized over the detection interval $(0,T)$, i.e., $\int dt' \, |\xi(t')|^2=1$. 
To analyze this measurement we build an orthonormal basis of temporal modes around the LO mode. This is physically motivated as the time dependence of $\xi_\lo$ is known and controlled in an experiment.

We construct this basis using the standard Gram-Schmidt process beginning with $\xi_\lo$ and $\xi_\sig$. We label this basis as $\{\xi_\lo, \xi_\perp, \xi_3,  \dots\}$\footnote{The numbered modes are necessary to complete the temporal mode basis, but will not contribute to the measurement operators. Additionally the above basis is ill-defined if $\xi_\lo \propto \xi_\sig$, but this is the mode-matched limit where theoretical treatments already exist.}, and define:
\begin{equation}\label{eqn:basis}
    \begin{aligned}
            \xi_\lo(t) &= \xi_\lo\\
            \xi_\perp(t) &= \frac{\xi_\sig - \<\xi_\lo,\xi_\sig\>\xi_\lo}{\sqrt{1-|\<\xi_\lo,\xi_\sig\>|^2}} %\\
            %&~~\vdots\\
    \end{aligned}
\end{equation}
where $\<f,g\> = \int_0^T dt' f^*(t')g(t')$ is the inner product. For convenience we also define the mode overlap, $\gamma \equiv \<\xi_\lo,\xi_\sig\>$. The measurement can be completely understood in these two modes because the signal can be decomposed into a linear combination of just $\xi_\lo$ and $\xi_\perp$. 

\begin{figure}[t]
    \centering
    \includegraphics[width=\columnwidth]{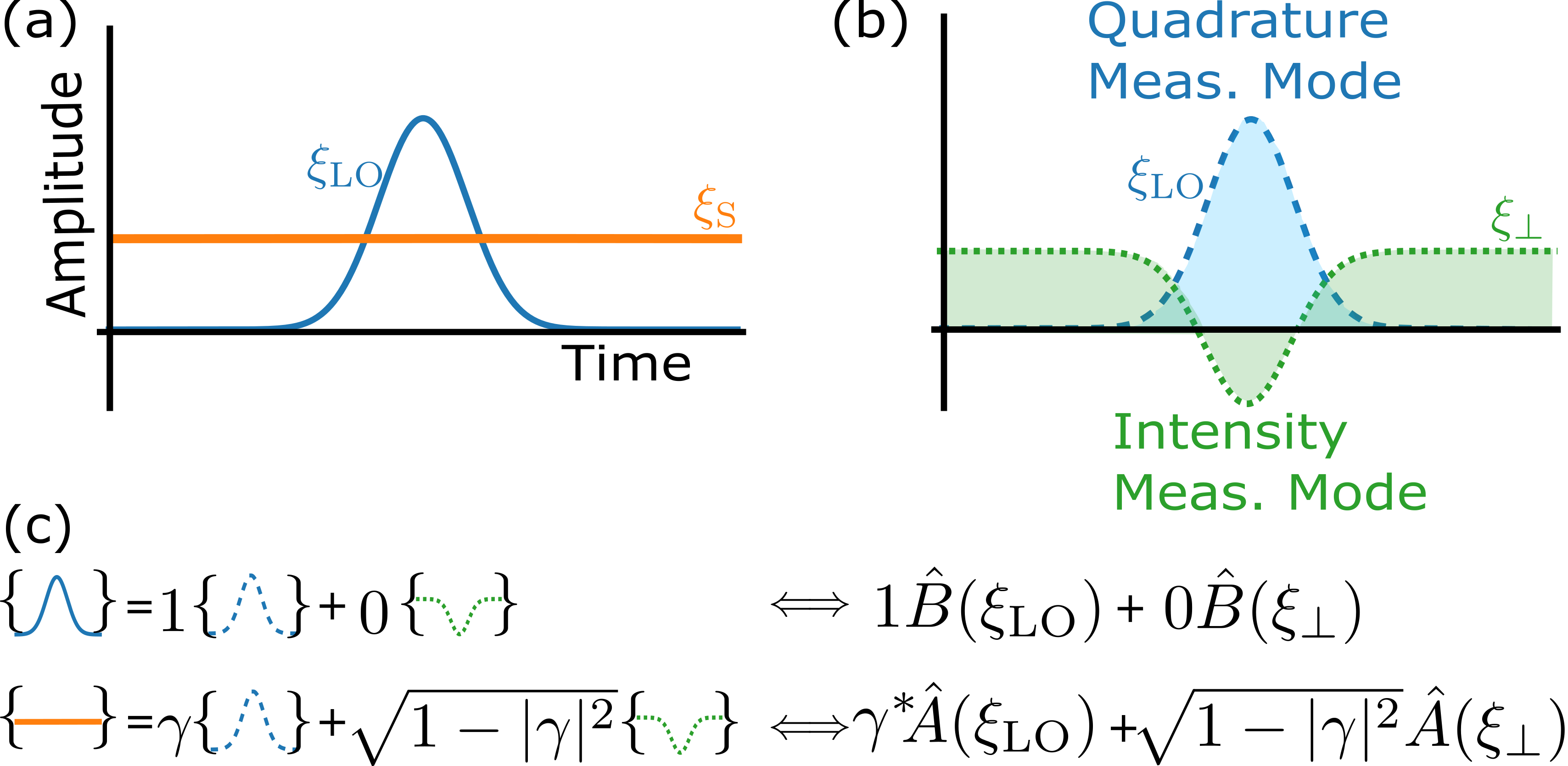}
    \caption{(a) The physical CW signal (orange) and Comb pulse LO (blue) modes incident on the beamsplitter. (b) The Gram-Schmidt modes which correspond to quadrature measurement and intensity-like measurement. (c) The signal and LO modes have some overlap, but we can decompose them into the Gram-Schmidt basis. We can write the signal and LO as a linear combination of these basis elements, here $\gamma$ is the complex-valued mode overlap, $\gamma = \<\xi_\lo,\xi_\sig\>$. This same decomposition can be used on the operators $\hat{B}(\xi_\lo)$ and $\hat{A}(\xi_\sig)$.}
    \label{fig:modes}
\end{figure}

To demonstrate how to decompose an example signal we can consider the case where the signal is continuous wave (CW) and the LO is a Gaussian pulse, representing, e.g., a short temporal section of a frequency comb. In the frame rotating at the carrier frequency we have,
\begin{equation}
    \xi_\sig = \frac{e^{i\phi}}{\sqrt{T}}, \quad {\rm and }\quad \xi_\lo = \left[\frac{e^{-(t-\mu)^2/(2\sigma^2)}}{\sqrt{2\pi\sigma^2}}\right ]^{1/2}.
\end{equation}
\cref{fig:modes} illustrates these modes with $\mu = T/2$, $\sigma \approx .1T$, and $\phi = 0$. Assuming the LO pulse is fully contained in the detection interval the mode overlap is $\gamma = \left(8\pi\sigma^2T^{-2}\right)^{1/4}e^{i\phi}$. The mode overlap is maximized when the ratio of the pulse width to the detection interval is maximal. Intuitively this is when the pulse is the most ``CW like" on the detection interval. Further the appearance of the phase $e^{i\phi}$ demonstrates that the mode overlap is complex in general. 

Using equation \cref{eqn:basis} we can also calculate the perpendicular mode
\begin{equation}
    \xi_\perp=  \frac{e^{i\phi}}{\sqrt{T - \sigma \sqrt{8\pi}}}\left(1- 2 e^{-(t-\mu)^2/(4\sigma^2)}\right),
\end{equation}
which is pictured in \cref{fig:modes}(b). This mode represents the piece of the signal that does not interact with the local oscillator. We will see in \cref{sec:modal_homodyne} that the perpendicular mode will contribute intensity like noise. 

Returning to the general case where the modes $\xi_\sig$ and $\xi_\lo$ are arbitrary, we define modal coherent states as ~\cite{Blow1990,loudon2000quantum, ou2017quantum},
\begin{align}
        |\alpha_{\xi_{\sig}}\> &= D(\alpha,\xi_{\sig})|0\> =\exp\left[\alpha \hat{A}^\dagger(\xi_{\sig}) - \alpha^*\hat{A}(\xi_{\sig}) \right]|0\>. 
\end{align}
It is straightforward to show that $A(\xi_\sig) = \gamma^*A(\xi_\lo)+\sqrt{1-|\gamma|^2}A(\xi_\perp)$. 
Using this along with an application of the Baker–Campbell–Hausdorff, and $[\hat{A}(\xi_1),\hat{A}^\dagger(\xi_2)] = \<\xi_1,\xi_2\>$ and we get the modal decomposition,
\begin{align}
    D(\alpha,\xi_\sig) = &D(\gamma\alpha,\xi_\lo) \otimes D(\sqrt{1-|\gamma|^2}\alpha,\xi_\perp)\nonumber\\
    &\times\exp\left\{|\alpha|^2\left(\text{Im}(\gamma^*\sqrt{1-|\gamma|^2}\<\xi_\perp,\xi_\lo\>\right)\right\}.
\end{align}
Since we have defined our mode basis to be orthonormal we know that $\<\xi_\perp,\xi_\lo\> = 0$ and thus the exponential term is 1. This allows us to decompose a coherent state signal into the LO and $\perp$ modes
\begin{equation}\label{eqn:cohdecomp}
\big|\alpha_{\xi_\sig}\big\>  = \big|\gamma\alpha_{\xi_\lo}\big\>\otimes\big|\sqrt{1-|\gamma|^2}\alpha_{\xi_\perp}\big\> .
\end{equation}
This will be useful in \cref{sec:examples} when we consider the measurement of a coherent signal. 

\section{Modal Homodyne Measurements}\label{sec:modal_homodyne}
The photodetectors in homodyne measurements respond to intensity. In the noiseless limit, intensity detection is photon number resolving. Like previous work~\cite{Tyc2004,Combes2022} we use photon number resolving detectors for our analysis, but the noisy detector limit can always be recovered by coarse graining.
However, in our analysis, we must consider photodetection in certain temporal modes.  
That is, we model photodetection as projections onto Fock states in a given mode: 
\begin{equation}\label{eqn:fock}
   \ket{n_\xi} = \frac{\hat{A}^\dagger(\xi)^n}{\sqrt{n!}}\ket{\vac} , 
\end{equation}
which are eigenstates of the number operator $\hat{A}^\dagger(\xi)\hat{A}(\xi)$. We use the notation where operators with modes are written with parentheses, $M_{n,m}(\xi)$, and states in modes are denoted with subscripts, $|n_\xi\rangle$.

We assume the detector is unable to differentiate a LO-mode photon from an orthogonal-mode photon.
  The detector acts as a projector onto a combination of all the possible modes in our basis that could produce a click. For this reason, we construct the $n$ click measurement operator by marginalizing over the mode degree of freedom 
\begin{equation}\label{eqn:detectorclick}
    |\vac \rangle\langle n|_D = \sum_{p=0}^n|\vac\rangle\left(\langle p_{\xi_\text{LO}}|\otimes\langle n-p_{\xi_\perp}|\right) \,,
\end{equation}
where we assume our detectors absorb photons, hence the projection onto vacuum.
Now we follow the analysis of Ref.~\cite{Combes2022} to arrive at the measurement (Kraus) operator that corresponds to observing $n$ clicks on one detector and $m$ clicks on the other:
\begin{equation}\label{eqn:twoportdetection}
    M_{n,m} = \<n|_{D1}\<m|_{D2} U_{\text{BS}} |\psi_\lo\>\, .
\end{equation}
We choose the LO to be a coherent state in $\xi_\lo$, $|\psi_\lo\> = |\beta({\xi_\lo})\>\otimes|0_{\xi_\perp}\>$. We assume our detectors absorb photons so we have used $\<n|_{D_1}$ as shorthand for $|\vac\>\<n|_{D_1}$. Here the operator $M_{n,m}$ is not given a mode because it pertains to the total clicks over all modes. At the moment these measurement operators are written in terms of $n$ and $m$, but ultimately we will express the POVM in terms of the difference and sum photocurrent $x \propto n-m$ and $w\propto n+m$ respectively.

We use the definition in \cref{eqn:detectorclick} and re-order the tensor product to write the measurement operators in our preferred basis
\begin{align}\label{eqn:convolution}
     M_{n,m} &= \sum_{p,q}\underbrace{\<p_{\xi_\lo}|\<q_{\xi_\lo}|}_{\lo\text{ modes}}
                \underbrace{\<n-p_{\xi_\perp}|\<m-q_{\xi_\perp}|}_{\perp\text{ modes}}U_{\text{BS}} |\psi_\lo\>\, ,\nonumber\\
            &= \sum_{p,q} M_{p,q}(\xi_\lo)\otimes M_{n-p,m-q}(\xi_{\perp})\,
\end{align}
where we assume the beamsplitter behaves uniformly across modes (\cref{app:Beamsplitter}),  although this assumption can be relaxed~\cite{avagyan_quantum_2023}.   \Cref{eqn:convolution} shows a decomposition of the measurement operators into two temporal modes
\begin{subequations}\label{eqn:seperated}
\begin{align}
    M_{p,q}(\xi_\lo) &= \<p_{\xi_\lo}|\<q_{\xi_\lo}|U_{\text{BS}}|\beta_{\xi_\lo}\>\label{eqn:LO_Quad}\\
    M_{r,s}(\xi_\perp) &= \<r_{\xi_\perp}|\<s_{\xi_\perp}|U_{\text{BS}}|\vac\>.\label{eqn:perp_Int}
\end{align}
\end{subequations}

To obtain our final result for this operator we need to leverage the methods of Refs.~\cite{Tyc2004,Combes2022} which involve four steps. {\em Step 1.}  We need to perform a change of variables on our operators so they are written in terms of the sum and difference variables. For both modes now we change from the $n$ and $m$ variables to sum and scaled difference variables
\begin{equation}\label{eqn:sumdiff}
x = \frac{n-m}{\sqrt{2}|\beta|e^{i\theta}} \quad \& \quad w = {n+m}, 
\end{equation}
where $\theta$ is the phase of the LO. We do this because we want our operators to describe the observed quantities of the measurement, which is the difference photocurrent. In the large LO limit, we will approximate $x$ as a continuous variable. 
{\em Step 2.} We construct the POVM from the measurement operators in \cref{eqn:convolution} as $E = M^\dagger M$ (\cref{app:POVM}). After a number of approximations we arrive at
\begin{equation}\label{eqn:POVMseperated}
    E_{x,w} = \sum_{w'}\int dx' E_{x',w'}(\xi_\lo)\otimes E_{x-x',w-w'}(\xi_\perp) \, , 
\end{equation}
where $E_{x,w}(\xi_\lo)$ and $E_{x,w}(\xi_\perp)$ are the POVMs for the measurement in each mode.
The POVM elements are the operators that gives rise to the statistics observed in an experiment.
{\em Step 3.} For the analysis of the LO mode we need to assume the LO is large so that we are in the homodyne limit, this entails assuming that $\<\hat{n}(\xi_\lo)\>_\sig \ll \<\hat{n}(\xi_\lo)\>_\lo $, i.e., the LO dominates the signal in the LO mode (\cref{app:largeLO}). Thus the difference variable $x$ is quasi-continuous.
 {\em Step 4.} We marginalize over the sum variable as it is not typically observed in experiments. We can now state our main result which is the total POVM for a time-dependent LO 
 \begin{equation}\label{eqn:POVMseperated137}
E_{x} = \sum_{w} E_{x,w}  = 
\int dx'  E_{x'}(\xi_\lo) \otimes  E_{x-x'}(\xi_{\perp}) \, ,
\end{equation}
this is a convolution of a POVM in the LO and perp modes ({\cref{app:convolution}}). We absorbed the Jacobian from changing variables into the single mode POVMs so that both the total and single mode POVMs sum to identity.  In \cref{eqn:POVMseperated137} $x$ on the LHS is the difference photocurrent while $x'$ and $x-x'\equiv v$ on the RHS are the difference variable contributions from the LO and $\perp$ mode respectively, c.f. \cref{eqn:convolution}. 

\begin{figure}[th!]
    \includegraphics[width=1\linewidth]{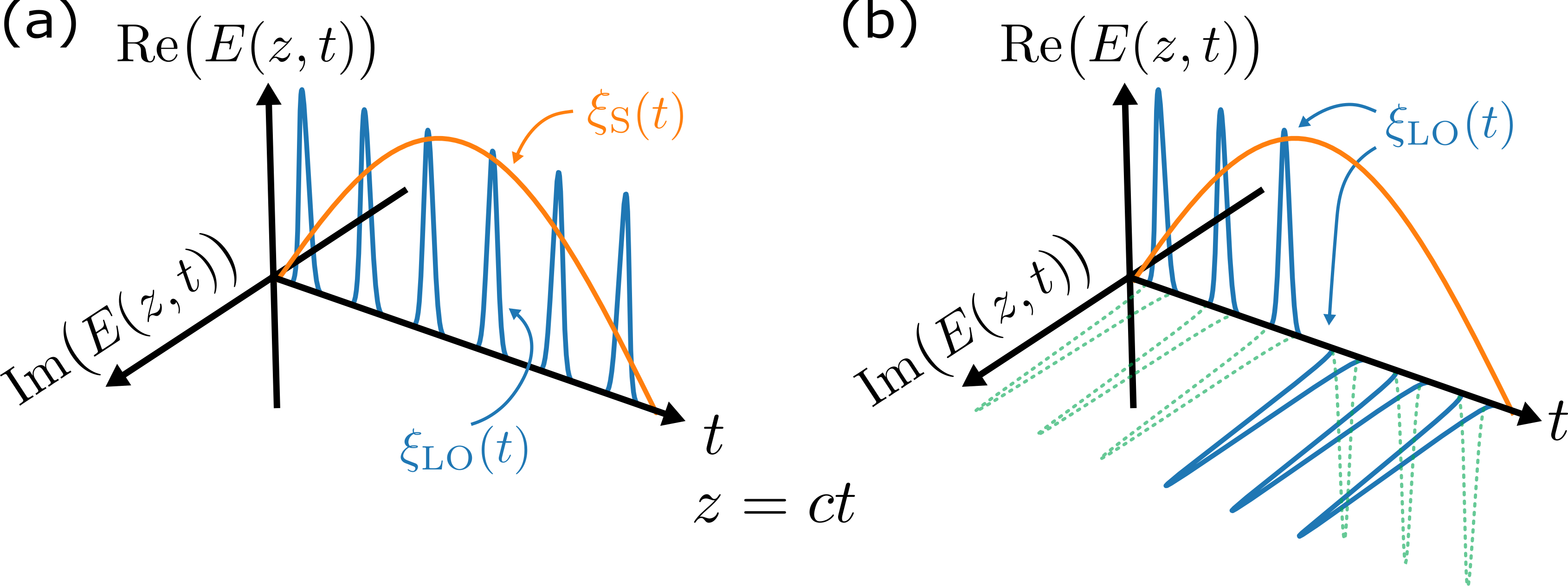} 
  \caption{The classical mode functions of a time dependant homodyne measurement. Here the mode  $\xi_\lo(t)$ is that of a comb, while the signal mode $\xi_\sig$ is half a period of a sinusoid. (a) the LO phase is constant over the signal and thus the modal quadrature measurement is consistent with a CW quadrature measurement. (b) the phase of the LO has an abrupt change so the time-dependent quadrature is inconsistent with CW quadrature measurement. Our formalism allows for such time-dependent quadratures. The orthogonal time-dependent quadrature $\hat P (\xi_\lo)= -i \big( \hat{A}^\dagger(\xi_\lo)- \hat{A}(\xi_\lo)\big)/\sqrt{2}$ is the dotted line.}
  \label{fig:time_dep_quad} 
\end{figure}

A detailed calculation  shows the LO mode POVM is a homodyne measurement of a time-dependent quadrature (\cref{app:POVM_LO})
 \begin{equation}\label{eqn:LOPOVM}
    E_{x'}(\xi_\lo) =|x'_{\xi_\lo}\>\<x_{\xi_\lo}'|,
\end{equation}
 here $|x'_{\xi_\lo}\>$ is an eigenstate of the modal quadrature operator $\hat{Q}(\xi_\lo) = \big(\hat{A}(\xi_\lo) + \hat{A}^\dagger(\xi_\lo)\big)/\sqrt{2}$. This quadrature operator is time-dependent in the sense that $\xi_\lo$ may have different phases at different times, see \cref{fig:time_dep_quad}.

In the perpendicular mode, the LO is in vacuum, and the analysis yields a POVM that follows a binomial distribution in $n$ and $m$. After switching to sum and difference variables we have~({\cref{app:POVM_perp}):
\begin{equation}\label{eqn:perpPOVM}
     E_{v}(\xi_{\perp}) =\sum_{w} |w_{\xi_\perp}\>\<w_{\xi_\perp}|\text{Bin}\Big(\frac{|\beta|v}{\sqrt{2}}\Big|w,\frac{1}{2},0\Big) \, ,
\end{equation}
where $\ket{w_{\xi_\perp}}$ is a Fock state and $\text{Bin}\big(x\big|n,p,\mu\big)$ is a binomial distribution characterised by $n$ and $p$ and shifted so that it has mean $\mu$ \cite{oh_maximum_2014}. Here we have used the difference variable $v$ as if it were discrete, but it must be approximated as continuous to be consistent with \cref{eqn:POVMseperated137}. In  \cref{eqn:perpPOVM} the difference variable distribution has mean 0 and variance $w/(2|\beta|^2)$, and this is true regardless of the signal state. Only the variance of the difference variable actually depends on the input state. In other words, this is not a quadrature measurement and instead resembles an intensity measurement, as evidenced by the projector onto Fock states.

We have derived the POVM starting from the slow detector limit. For detectors that can resolve the time dependence of the signal and LO the Kraus operators would change. This time-dependent photo-record must be treated carefully. So long as the additional time dependence is averaged over, the measurement statistics predicted by the POVM presented here would remain correct, meaning our POVM is valid in any detector bandwidth limit ({\cref{app:time_dep}}). 

%========================================
\section{Examples}\label{sec:examples}
%========================================
We now illustrate the use of these theoretical tools with three examples. First, we apply our results to coherent signals and explore the limits of filtering. In the first two examples we apply filtering to a signal with known and unknown temporal profiles. These examples explain a remarkable demonstration by \citet{Deschenes2013} of higher SNR than that set by shot noise of the total photocurrent, achieved by filtering the measurement record. Finally, we utilize our measurement operators to analyze the measurement of a single photon which cannot be calculated using semi-classical methods. Moreover, the filtering technique we explore could be of interest to weak field homodyne when there is mismatch between the weak LO and the signal~
\cite{Vogel1995,ThekkadathWalmsley2020,OlivaresBondani2020}.

\subsection{Coherent state signal and SNR bound} \label{sec:example1}

We take the signal to be a coherent state $|\psi\>_\sig = |\alpha_{\xi_\sig}\> $ and  decompose the signal into the LO mode and the perpendicular ($\perp$) mode as in \cref{eqn:cohdecomp}.

We derive the distribution of the total measurement by taking the expectation of the POVM, \cref{eqn:POVMseperated137}, which is $ P(x) = \<\alpha_{\xi_\sig}|E_x|\alpha _{\xi_\sig}\>$ ({\cref{app:x_var}}). If the detector could differentiate photons in the perpendicular and LO modes, then the joint distribution of clicks in the two modes would be 
\begin{equation}\label{eqn:jointdist}
\begin{aligned}
P(x_\lo,x_\perp) = \mathcal{N}\left(x_\lo|\mu_\lo, \sigma_\lo^2 \right)\mathcal{N}\left(x_\perp|\mu_\perp,\sigma_\perp^2\right) ,
\end{aligned}
\end{equation}
where $\mathcal{N}$ denotes a normal distribution, $\mu_\lo = \sqrt{2}\text{Re}(\alpha\gamma)$, $\sigma_\lo^2 = 1/2$, $\mu_\perp= 0$,  and $\sigma_\perp^2 = |\alpha|^2(1-|\gamma|^2)/2|\beta|^2 $.
Here we have used the convention that $\beta$ is real, i.e., the phase of LO is the reference phase. Because the signal is a coherent state, which is uncorrelated in time, the distributions of $x_\perp$ and $x_\lo$ are independent. %This is not true for all input states. 
This means if we marginalize, or filter, over the $\perp$-mode we could obtain an ``ideal'' homodyne measurement of the signal in the LO mode.

The real detectors cannot differentiate between the two modes so the distributions must be convolved yielding,
\begin{equation}\label{eqn:total_dist}
   P(x)= \mathcal{N}\left(\mu=\sqrt{2}\text{Re}(\gamma\alpha),\sigma^2 = \frac{1}{2}+\frac{|\alpha|^2(1-|\gamma|^2)}{2|\beta|^2}\right).
\end{equation}
From this, we find the power signal-to-noise ratio (SNR):
\begin{equation}
    \text{SNR} = \frac{|\mu|^2}{\sigma^2} = \frac{4|\beta|^2\text{Re}(\gamma\alpha)^2}{|\beta|^2+(1-|\gamma|^2)|\alpha|^2}.
\end{equation}
As the signal and LO become mode-matched, i.e., $|\gamma|^2\to 1$, we recover the expected SNR for ideal homodyne detection: $\text{SNR}_{|\gamma|\to 1} = 4\text{Re}(\gamma\alpha)^2$. 

 When $\gamma$ is small and $|\beta|^2\gg |\gamma\alpha|^2$, i.e., large mode-mismatch, a Taylor expansion yields 
 \begin{equation}\label{eqn:SNR_comp}
 \begin{aligned}
\text{SNR}_{\gamma\to 0}^{\text{total}} \approx \frac{4|\beta|^2\text{Re}(\gamma\alpha)^2 }{|\beta|^2 +|\alpha|^2}\,. 
 \end{aligned}
 \end{equation} 
Thus ${\rm SNR}^{\rm total}$ has its mean attenuated by the mode-mismatch $\text{Re}(\gamma)$ as predicted by prior theory~\cite{Grosshans2001,Polycarpou2012,Qin2015,ChenWangYu2023}. Additionally, there are shot noise contributions from both the signal and LO since no assumption allows either noise term to dominate. The above SNR is conventionally accepted as the quantum limit for frequency comb measurements \cite{REICHERT1999}.

\begin{figure}[th!]
    \includegraphics[width=1\linewidth]{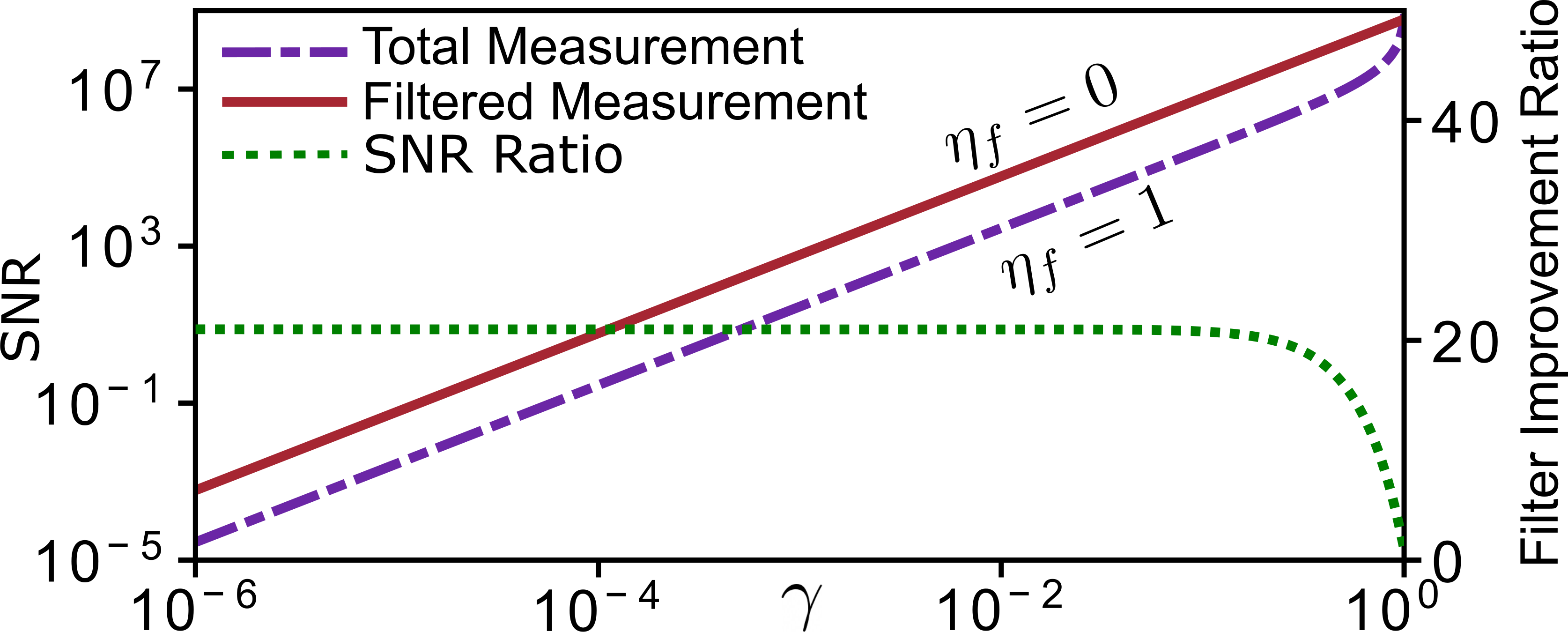} 
  \caption{\ SNR vs the mode overlap $\gamma$ with an experimentally relevant set of parameters. The dashed line is the total measurement SNR containing both LO mode and perpendicular mode clicks ($\eta_{f}=1$). The solid line is the SNR after an optimal filter is applied ($\eta_{f}=0$). The dotted line is the improvement obtainable using filtering. For this example we take the signal power to be 2 mW, the LO power is 100 $\mu$W, and the measurement interval is $\tau = 10$ ns $ = 1/f_{rep}$. For a 10 ps pulse and CW signal we would have typical $\gamma$ of $10^{-2}$ and $\eta_{f}$ of $10^{-3}$; here the available SNR gain is  $\approx 13$  dB.}
  \label{fig:results} 
\end{figure}

This conventional SNR limit was first questioned in an experiment by Deschênes and Genest \cite{Deschenes2013}, where they applied a filter matched to the LO intensity and achieved a sizeable SNR improvement over \cref{eqn:SNR_comp}. We reconsider this technique specifically for the case where we want to measure the time dependent quadrature operator $\hat{Q}(\xi_\lo)$ and the mode of the signal $\xi_\sig$ is unknown.
 
The optimal filter of the photocurrent is described by the time-dependent weighting function,
\begin{equation}
    f(t) = \begin{cases} 
      1 & \text{if } \xi_{\lo}(t) \ne 0  \\
      0 & \text{if } \xi_\lo(t) = 0
   \end{cases}\, ,
\end{equation}
which must be approximated in many cases {\red (\app I)}. This function leaves the LO mode unchanged,  so it will preserve the mode of the measured quadrature and leave the mean, unchanged. This filter will reduce the shot noise contribution from the perpendicular mode. The achievable filtered SNR is
\begin{equation}
    \text{SNR}^{f(t)} = \frac{4|\beta|^2\text{Re}(\gamma\alpha)^2}{|\beta|^2 + \eta_{f}|\alpha|^2},
\end{equation}
where $\eta_{f}$ is the filtering inefficiency given by $\eta_{f} = \int dt|f(t)|^2|\xi_\perp(t)|^2$. When $\eta_{f} = 0$ (perfect filtering) we recover the ideal homodyne SNR and when $\eta_{f} = 1$ (no filtering) we have the conventional SNR limit; i.e.
\begin{equation}\label{eq:badabing}
    \underbrace{\frac{4|\beta|^2\text{Re}(\gamma\alpha)^2 }{|\beta|^2 +|\alpha|^2}}_{\text{Conventional}}\le \,\underbrace{\frac{4|\beta|^2\text{Re}(\gamma\alpha)^2}{|\beta|^2 + \eta_{f}|\alpha|^2}}_{\text{Achievable}}\le \underbrace{\!\!\phantom{\frac{|}{|^1}}\!\!4\text{Re}(\gamma\alpha)^2}_{\text{Ideal}} \, .
\end{equation}
We plot these limits for parameter values typical in comb experiments in \cref{fig:results} and show a 13 dB improvement from perfect filtering. Note that in this derivation, the ``comb'' structure of the LO is, by definition, included in the mode $\xi_\lo$ (see Fig. \ref{fig:time_dep_quad}), and its impact is described by the overlap with the signal mode $\xi_\sig$, i.e., the parameter $\gamma$.

We caution that the SNR bounds in \cref{eq:badabing} can be beaten if there is prior information about the shape of the signal mode. However, doing so will cause the measurement to no longer be of $\hat{Q}(\xi_\lo)$. For example, if the signal mode is zero while the LO mode is nonzero, that portion of the measurement record only contributes LO shot noise and does not change the mean and could thus be ignored to increase SNR, as explained in \cref{app:square_signal}.

% ========================================================
\subsection{A known signal mode can violate SNR bound}\label{app:square_signal}
% ========================================================
Now we consider an example where the SNR bounds we proposed can be violated by changing the effective quadrature being measured. We take the signal and LO modes to both be top hat functions, 
\begin{equation}
    \xi_\sig =\begin{cases} 
      \frac{1}{\sqrt{t_1}}& \text{if } t\in (0,t_1)  \\
      0 & \text{else}
   \end{cases}~~~~\xi_\lo =  \begin{cases} 
      \frac{1}{\sqrt{\tau}} & \text{if } t\in (t_0,T)  \\
      0 & \text{else}
   \end{cases},
\end{equation}
where $\tau =T-t_0$, see \cref{fig:tophat}. The LO defines the modal measurement. As the signal and LO only overlap on the interval $(t_0,t_1)$ we are predominantly measuring vacuum signal.

\begin{figure}[th!]
\centering
    \includegraphics[width=1\columnwidth]{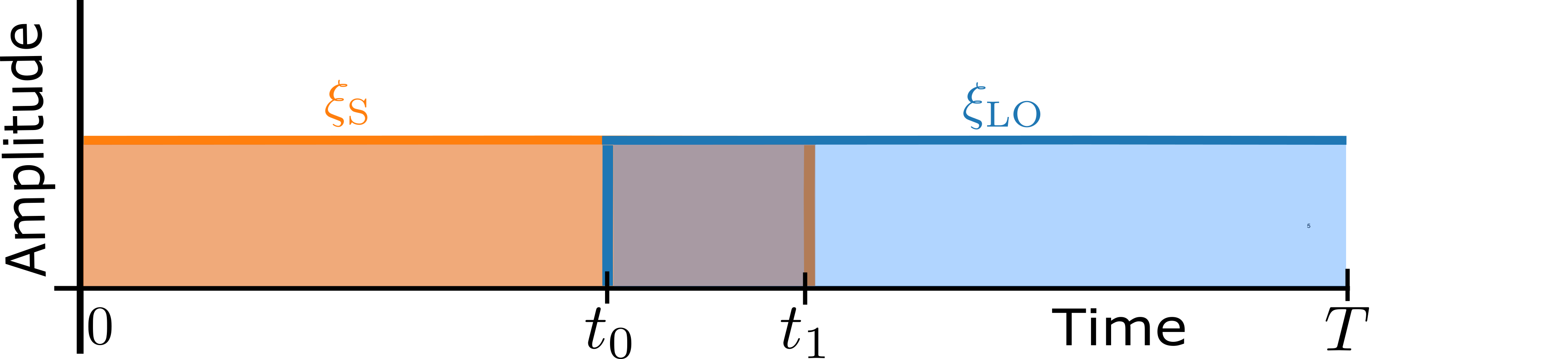} 
  \caption{We show the modes used to demonstrate filtering performance that exceeds our bounds. Here it is clear that the optimal gate would be the one that only count clicks from the overlapped portion of the measurement and remove all additional shot noise from the LO and signal mode.}
  \label{fig:tophat} 
\end{figure}

As the signal mode function is known we may filter out the measurement of vacuum by only considering detection in the interval $(t_0,t_1)$. This is achieved with the filter
\begin{equation}
    f(t) = \begin{cases} 
      1& \text{if } t\in (t_0,t_1)  \\
      0 & \text{else}
      \end{cases}.
\end{equation}
This will change the quadrature measured from $\hat Q(\xi_{\lo})$ to the top hat ``LO mode'' in the interval $\hat Q(f(t)\xi_\lo)$.

Redoing the above analysis in this case for a coherent signal $|\alpha_{\xi_\sig}\>$, the SNR is
\begin{equation}
    \text{SNR} = \frac{4\text{Re}(\gamma\alpha\beta^*)^2}{\eta_\lo|\beta|^2 + \eta_\sig|\alpha|^2},
\end{equation}
where $\eta_\lo = (t_1-t_0)/(T-t_0)$, and  $\eta_\sig = (t_1-t_0)/t_1$. In the large LO regime after the filter is applied we get the simplified expression
\begin{equation}
    \text{SNR} = \frac{4\text{Re}(\gamma\alpha)^2}{\eta_\lo},
\end{equation}
where we note that $\eta_\lo$ is between 0 and 1, so this exceeds the bound we propose of $\text{SNR} = 4\text{Re}(\gamma\alpha)^2$ by a factor of $\eta_\lo^{-1}$. Thus in this example, because we knew $\xi_\sig$, we were able to violate the SNR bound at the cost of altering the modal measurement.

\subsection{Single photon signal and weak LO}

Now consider the signal to be a single photon state with mode $\xi_s$. Decomposing this into our basis gives
\begin{equation}\label{eqn:singlephoton}
    |1_{\xi_s}\> = \gamma |1_{\xi_\lo}\>|0_{\xi_\perp} \>+\sqrt{1-|\gamma|^2}|0_{\xi_\lo}\>|1_{\xi_\perp}\> \,. 
\end{equation}
As the overlap between the signal and LO modes increases the photon is more likely to be found in the LO mode.

To compute the measured quadrature distribution we take the expectation of the POVM in \cref{eqn:POVMseperated137}  in the state \cref{eqn:singlephoton} ie.
\begin{equation}
P(x) =    \int dx' \<1_{\xi_s}| E_{x'}^\lo(\xi_\lo) \otimes  E_{x-x'}^\perp(\xi_{\perp})|1_{\xi_s}\> \, .
\end{equation}
After some manipulation done in \cref{app:1photon} we find 

\begin{equation}\label{eqn:single_dist}
\begin{aligned}
     P(x) = \frac{1}{2\sqrt{\pi}} \Big [ & 4|\gamma|^2 x^2 e^{-x^2} + (1-|\gamma|^2)\times\\ &(e^{-(x-1/\sqrt{2}|\beta|)^2}+e^{-(x+1/\sqrt{2}|\beta|)^2})\Big] \, .
\end{aligned}
\end{equation}
The first term is the quadrature distribution for a single photon which is attained in the mode-matched limit $(\gamma = 1)$. For $\gamma = 0$ and $|\beta|\gg 1$ we have the quadrature distribution for vacuum. For $\gamma$ between these extremes and large $\beta$ we have the quadrature distribution for a mixed state of zero and one photons. This is the regime that is characterized by an effective loss and is analyzed in detail in \cite{Grosshans2001}. For small values of $\beta$, additional shot noise from the perpendicular mode splits the Gaussian distribution of vacuum into two Gaussians. These features are plotted in figure \cref{fig:single}. The Filtered measurement is equivalent to the effective loss results from \cite{Grosshans2001}, showcasing the difference the added perpendicular mode noise can make. This feature is not present if mode mismatch is treated as just loss and is highly relevant to applications of homodyne with weak LOs \cite{Vogel1995,ThekkadathWalmsley2020,OlivaresBondani2020}.

\begin{figure}[!th]
    \includegraphics[width=1\linewidth]{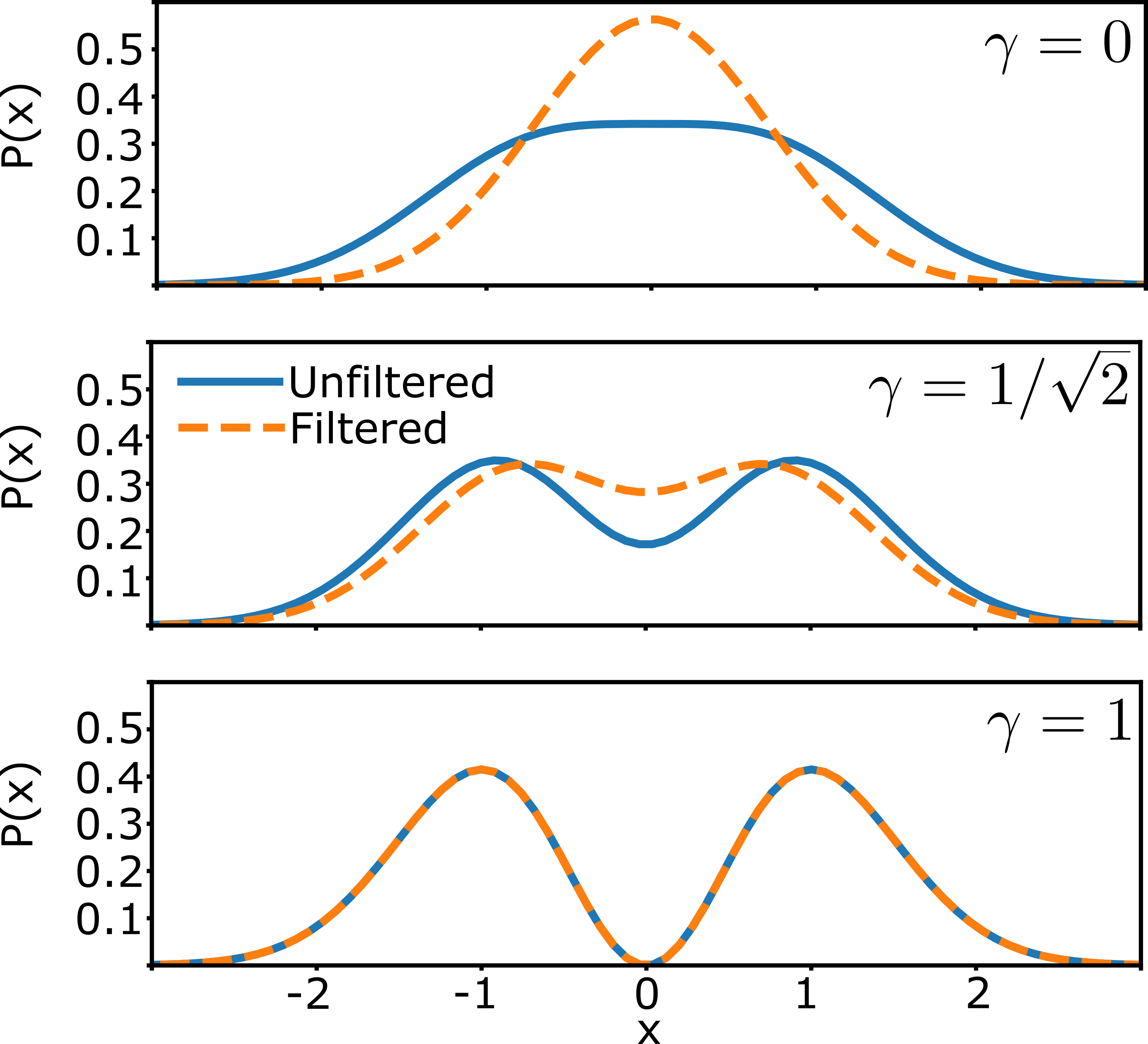} 
  \caption{Difference variable probability distribution for a single photon signal with varying amounts of mode mismatch. In all figures $\beta = 1$ is used, so that the single photon of shot noise from the signal is not overshadowed. These figures are generated with $\eta_f = 0 $. Qualitatively similar results can be observed for larger photon number signal states, but combinatorial expansion complicates the analytical formulations equivalent to \cref{eqn:single_dist}.}
  \label{fig:single} 
\end{figure}

% ======================
\section{conclusion}\label{sec:conclusion}
% ======================
In conclusion, we have found the measurement operators for multimode homodyne detection that are valid for arbitrary time dependence of both the LO and signal. In our construction, the measurement decomposes naturally into two parts: a quadrature measurement in the temporal mode of the LO and an intensity-like measurement on the other modes. 
We show that perfect filtering of this intensity noise achieves a quadrature noise-limited measurement. In comb-based measurements characterized by large mode-mismatch and strong signals, this establishes a significantly lower quantum limit than conventionally considered. This limit should be sought before pursuing quantum advantage from non-classical states of light. Moreover, because we have developed a fully quantum theory, we can analyze the measurement of any signal state including squeezed states, which are highly relevant to measurements aimed at increased precision. As an example of our fully quantum theory, we analyzed the measurement of a single photon Fock state with a finite strength LO and arbitrary mode overlap. This represents a scenario that existing methods have not been able to fully describe. Our analysis complements related work on POVMs for electro-optic sampling~\cite{HubenschmidBurkard2022}

 Many comb-based measurements are based on heterodyne rather than homodyne techniques. Although there are many reasons to prefer homodyne over heterodyne for quantum metrology, the technical limitations of frequency comb measurements make quantum-limited homodyne more difficult to achieve than heterodyne.
 We conjecture that the heterodyne SNR is simply half that of homodyne due to sampling of both $\hat x$ and $\hat p$ quadratures, but a complete analysis of the heterodyne measurement operators is left as future work.

{\em Acknowledgments:} We wish to acknowledge helpful discussions with Joseph Bush, Jérôme Genest, Scott Glancy, Matthew Heyrich, Molly Kate Kreider, William Schenken, Takeshi Umeki, and Mathieu Walsh. This work was supported by the NSF through the QLCI award OMA-2016244 and by ONR award N00014-21-1-2606. E.J.T. acknowledges funding support from the NDSEG Fellowship

\bibliography{references}

%Control: production of eprint (1) enabled
\begin{thebibliography}{50}%
\makeatletter
\providecommand \@ifxundefined [1]{%
 \@ifx{#1\undefined}
}%
\providecommand \@ifnum [1]{%
 \ifnum #1\expandafter \@firstoftwo
 \else \expandafter \@secondoftwo
 \fi
}%
\providecommand \@ifx [1]{%
 \ifx #1\expandafter \@firstoftwo
 \else \expandafter \@secondoftwo
 \fi
}%
\providecommand \natexlab [1]{#1}%
\providecommand \enquote  [1]{``#1''}%
\providecommand \bibnamefont  [1]{#1}%
\providecommand \bibfnamefont [1]{#1}%
\providecommand \citenamefont [1]{#1}%
\providecommand \href@noop [0]{\@secondoftwo}%
\providecommand \href [0]{\begingroup \@sanitize@url \@href}%
\providecommand \@href[1]{\@@startlink{#1}\@@href}%
\providecommand \@@href[1]{\endgroup#1\@@endlink}%
\providecommand \@sanitize@url [0]{\catcode `\\12\catcode `\$12\catcode
  `\&12\catcode `\#12\catcode `\^12\catcode `\_12\catcode `\%12\relax}%
\providecommand \@@startlink[1]{}%
\providecommand \@@endlink[0]{}%
\providecommand \url  [0]{\begingroup\@sanitize@url \@url }%
\providecommand \@url [1]{\endgroup\@href {#1}{\urlprefix }}%
\providecommand \urlprefix  [0]{URL }%
\providecommand \Eprint [0]{\href }%
\providecommand \doibase [0]{https://doi.org/}%
\providecommand \selectlanguage [0]{\@gobble}%
\providecommand \bibinfo  [0]{\@secondoftwo}%
\providecommand \bibfield  [0]{\@secondoftwo}%
\providecommand \translation [1]{[#1]}%
\providecommand \BibitemOpen [0]{}%
\providecommand \bibitemStop [0]{}%
\providecommand \bibitemNoStop [0]{.\EOS\space}%
\providecommand \EOS [0]{\spacefactor3000\relax}%
\providecommand \BibitemShut  [1]{\csname bibitem#1\endcsname}%
\let\auto@bib@innerbib\@empty
%</preamble>
\bibitem [{\citenamefont {Kumar}\ \emph {et~al.}(2012)\citenamefont {Kumar},
  \citenamefont {Barrios}, \citenamefont {MacRae}, \citenamefont {Cairns},
  \citenamefont {Huntington},\ and\ \citenamefont
  {Lvovsky}}]{HuntingtonLovovsky2012}%
  \BibitemOpen
  \bibfield  {author} {\bibinfo {author} {\bibfnamefont {R.}~\bibnamefont
  {Kumar}}, \bibinfo {author} {\bibfnamefont {E.}~\bibnamefont {Barrios}},
  \bibinfo {author} {\bibfnamefont {A.}~\bibnamefont {MacRae}}, \bibinfo
  {author} {\bibfnamefont {E.}~\bibnamefont {Cairns}}, \bibinfo {author}
  {\bibfnamefont {E.~H.}\ \bibnamefont {Huntington}},\ and\ \bibinfo {author}
  {\bibfnamefont {A.~I.}\ \bibnamefont {Lvovsky}},\ }\bibfield  {title}
  {\bibinfo {title} {Versatile wideband balanced detector for quantum optical
  homodyne tomography},\ }\href
  {https://doi.org/https://doi.org/10.1016/j.optcom.2012.07.103} {\bibfield
  {journal} {\bibinfo  {journal} {Optics Communications}\ }\textbf {\bibinfo
  {volume} {285}},\ \bibinfo {pages} {5259} (\bibinfo {year}
  {2012})}\BibitemShut {NoStop}%
\bibitem [{\citenamefont {Gerrits}\ \emph {et~al.}(2011)\citenamefont
  {Gerrits}, \citenamefont {Glancy},\ and\ \citenamefont {Nam}}]{Gerrits2011}%
  \BibitemOpen
  \bibfield  {author} {\bibinfo {author} {\bibfnamefont {T.}~\bibnamefont
  {Gerrits}}, \bibinfo {author} {\bibfnamefont {S.}~\bibnamefont {Glancy}},\
  and\ \bibinfo {author} {\bibfnamefont {S.~W.}\ \bibnamefont {Nam}},\
  }\bibfield  {title} {\bibinfo {title} {{A balanced homodyne detector and
  local oscillator shaping for measuring optical Schr{\"{o}}dinger cat
  states}},\ }\href {https://doi.org/10.1117/12.885486} {\bibfield  {journal}
  {\bibinfo  {journal} {Advanced Photon Counting Techniques V}\ }\textbf
  {\bibinfo {volume} {8033}},\ \bibinfo {pages} {80330X} (\bibinfo {year}
  {2011})}\BibitemShut {NoStop}%
\bibitem [{\citenamefont {Fuwa}\ \emph {et~al.}(2015)\citenamefont {Fuwa},
  \citenamefont {Takeda}, \citenamefont {Zwierz}, \citenamefont {Wiseman},\
  and\ \citenamefont {Furusawa}}]{Fuwa2015}%
  \BibitemOpen
  \bibfield  {author} {\bibinfo {author} {\bibfnamefont {M.}~\bibnamefont
  {Fuwa}}, \bibinfo {author} {\bibfnamefont {S.}~\bibnamefont {Takeda}},
  \bibinfo {author} {\bibfnamefont {M.}~\bibnamefont {Zwierz}}, \bibinfo
  {author} {\bibfnamefont {H.~M.}\ \bibnamefont {Wiseman}},\ and\ \bibinfo
  {author} {\bibfnamefont {A.}~\bibnamefont {Furusawa}},\ }\bibfield  {title}
  {\bibinfo {title} {{Experimental proof of nonlocal wavefunction collapse for
  a single particle using homodyne measurements}},\ }\href
  {https://doi.org/10.1038/ncomms7665} {\bibfield  {journal} {\bibinfo
  {journal} {Nature Communications}\ }\textbf {\bibinfo {volume} {6}},\
  \bibinfo {pages} {1} (\bibinfo {year} {2015})}\BibitemShut {NoStop}%
\bibitem [{\citenamefont {Polycarpou}\ \emph {et~al.}(2012)\citenamefont
  {Polycarpou}, \citenamefont {Cassemiro}, \citenamefont {Venturi},
  \citenamefont {Zavatta},\ and\ \citenamefont {Bellini}}]{Polycarpou2012}%
  \BibitemOpen
  \bibfield  {author} {\bibinfo {author} {\bibfnamefont {C.}~\bibnamefont
  {Polycarpou}}, \bibinfo {author} {\bibfnamefont {K.~N.}\ \bibnamefont
  {Cassemiro}}, \bibinfo {author} {\bibfnamefont {G.}~\bibnamefont {Venturi}},
  \bibinfo {author} {\bibfnamefont {A.}~\bibnamefont {Zavatta}},\ and\ \bibinfo
  {author} {\bibfnamefont {M.}~\bibnamefont {Bellini}},\ }\bibfield  {title}
  {\bibinfo {title} {{Adaptive detection of arbitrarily shaped ultrashort
  quantum light states}},\ }\href
  {https://dx.doi.org/10.1103/PhysRevLett.109.053602} {\bibfield  {journal}
  {\bibinfo  {journal} {Physical Review Letters}\ }\textbf {\bibinfo {volume}
  {109}} (\bibinfo {year} {2012})}\BibitemShut {NoStop}%
\bibitem [{\citenamefont {Qin}\ \emph {et~al.}(2015)\citenamefont {Qin},
  \citenamefont {Prasad}, \citenamefont {Brannan}, \citenamefont {MacRae},
  \citenamefont {Lezama},\ and\ \citenamefont {Lvovsky}}]{Qin2015}%
  \BibitemOpen
  \bibfield  {author} {\bibinfo {author} {\bibfnamefont {Z.}~\bibnamefont
  {Qin}}, \bibinfo {author} {\bibfnamefont {A.~S.}\ \bibnamefont {Prasad}},
  \bibinfo {author} {\bibfnamefont {T.}~\bibnamefont {Brannan}}, \bibinfo
  {author} {\bibfnamefont {A.}~\bibnamefont {MacRae}}, \bibinfo {author}
  {\bibfnamefont {A.}~\bibnamefont {Lezama}},\ and\ \bibinfo {author}
  {\bibfnamefont {A.~I.}\ \bibnamefont {Lvovsky}},\ }\bibfield  {title}
  {\bibinfo {title} {{Complete temporal characterization of a single photon}},\
  }\href {https://doi.org/10.1038/lsa.2015.71} {\bibfield  {journal} {\bibinfo
  {journal} {Light: Science and Applications}\ }\textbf {\bibinfo {volume}
  {4}},\ \bibinfo {pages} {e298} (\bibinfo {year} {2015})}\BibitemShut
  {NoStop}%
\bibitem [{\citenamefont {Grandi}\ \emph {et~al.}(2017)\citenamefont {Grandi},
  \citenamefont {Zavatta}, \citenamefont {Bellini},\ and\ \citenamefont
  {Paris}}]{GrandiParis2017}%
  \BibitemOpen
  \bibfield  {author} {\bibinfo {author} {\bibfnamefont {S.}~\bibnamefont
  {Grandi}}, \bibinfo {author} {\bibfnamefont {A.}~\bibnamefont {Zavatta}},
  \bibinfo {author} {\bibfnamefont {M.}~\bibnamefont {Bellini}},\ and\ \bibinfo
  {author} {\bibfnamefont {M.~G.~A.}\ \bibnamefont {Paris}},\ }\bibfield
  {title} {\bibinfo {title} {Experimental quantum tomography of a homodyne
  detector},\ }\href {https://doi.org/10.1088/1367-2630/aa6f2c} {\bibfield
  {journal} {\bibinfo  {journal} {New Journal of Physics}\ }\textbf {\bibinfo
  {volume} {19}},\ \bibinfo {pages} {053015} (\bibinfo {year}
  {2017})}\BibitemShut {NoStop}%
\bibitem [{\citenamefont {Walker}(1987)}]{Walker87}%
  \BibitemOpen
  \bibfield  {author} {\bibinfo {author} {\bibfnamefont {N.}~\bibnamefont
  {Walker}},\ }\bibfield  {title} {\bibinfo {title} {Quantum theory of
  multiport optical homodyning},\ }\href
  {https://doi.org/10.1080/09500348714550131} {\bibfield  {journal} {\bibinfo
  {journal} {Journal of Modern Optics}\ }\textbf {\bibinfo {volume} {34}},\
  \bibinfo {pages} {15} (\bibinfo {year} {1987})}\BibitemShut {NoStop}%
\bibitem [{\citenamefont {Braunstein}(1990)}]{Braunstein90}%
  \BibitemOpen
  \bibfield  {author} {\bibinfo {author} {\bibfnamefont {S.~L.}\ \bibnamefont
  {Braunstein}},\ }\bibfield  {title} {\bibinfo {title} {Homodyne statistics},\
  }\href {https://doi.org/10.1103/PhysRevA.42.474} {\bibfield  {journal}
  {\bibinfo  {journal} {Phys. Rev. A}\ }\textbf {\bibinfo {volume} {42}},\
  \bibinfo {pages} {474} (\bibinfo {year} {1990})}\BibitemShut {NoStop}%
\bibitem [{\citenamefont {Banaszek}\ and\ \citenamefont
  {W\'odkiewicz}(1997)}]{BanaszekWodkiewicz1997}%
  \BibitemOpen
  \bibfield  {author} {\bibinfo {author} {\bibfnamefont {K.}~\bibnamefont
  {Banaszek}}\ and\ \bibinfo {author} {\bibfnamefont {K.}~\bibnamefont
  {W\'odkiewicz}},\ }\bibfield  {title} {\bibinfo {title} {Operational theory
  of homodyne detection},\ }\href {https://doi.org/10.1103/PhysRevA.55.3117}
  {\bibfield  {journal} {\bibinfo  {journal} {Phys. Rev. A}\ }\textbf {\bibinfo
  {volume} {55}},\ \bibinfo {pages} {3117} (\bibinfo {year}
  {1997})}\BibitemShut {NoStop}%
\bibitem [{\citenamefont {Tyc}\ and\ \citenamefont {Sanders}(2004)}]{Tyc2004}%
  \BibitemOpen
  \bibfield  {author} {\bibinfo {author} {\bibfnamefont {T.}~\bibnamefont
  {Tyc}}\ and\ \bibinfo {author} {\bibfnamefont {B.~C.}\ \bibnamefont
  {Sanders}},\ }\bibfield  {title} {\bibinfo {title} {{Operational formulation
  of homodyne detection}},\ }\href
  {https://doi.org/10.1088/0305-4470/37/29/010} {\bibfield  {journal} {\bibinfo
   {journal} {Journal of Physics A: Mathematical and General}\ }\textbf
  {\bibinfo {volume} {37}},\ \bibinfo {pages} {7341} (\bibinfo {year}
  {2004})}\BibitemShut {NoStop}%
\bibitem [{\citenamefont {Helstrom}(1967)}]{Helstrom1967}%
  \BibitemOpen
  \bibfield  {author} {\bibinfo {author} {\bibfnamefont {C.~W.}\ \bibnamefont
  {Helstrom}},\ }\bibfield  {title} {\bibinfo {title} {{Detectability of
  Coherent Optical Signals in a Heterodyne Receiver}},\ }\href
  {https://doi.org/10.1364/josa.57.000353} {\bibfield  {journal} {\bibinfo
  {journal} {Journal of the Optical Society of America}\ }\textbf {\bibinfo
  {volume} {57}},\ \bibinfo {pages} {353} (\bibinfo {year} {1967})}\BibitemShut
  {NoStop}%
\bibitem [{\citenamefont {Shapiro}(1985)}]{Shapiro85}%
  \BibitemOpen
  \bibfield  {author} {\bibinfo {author} {\bibfnamefont {J.}~\bibnamefont
  {Shapiro}},\ }\bibfield  {title} {\bibinfo {title} {Quantum noise and excess
  noise in optical homodyne and heterodyne receivers},\ }\href
  {https://doi.org/10.1109/JQE.1985.1072640} {\bibfield  {journal} {\bibinfo
  {journal} {IEEE Journal of Quantum Electronics}\ }\textbf {\bibinfo {volume}
  {21}},\ \bibinfo {pages} {237} (\bibinfo {year} {1985})}\BibitemShut
  {NoStop}%
\bibitem [{\citenamefont {Collett}\ \emph {et~al.}(1987)\citenamefont
  {Collett}, \citenamefont {Loudon},\ and\ \citenamefont
  {Gardiner}}]{Collett87}%
  \BibitemOpen
  \bibfield  {author} {\bibinfo {author} {\bibfnamefont {M.}~\bibnamefont
  {Collett}}, \bibinfo {author} {\bibfnamefont {R.}~\bibnamefont {Loudon}},\
  and\ \bibinfo {author} {\bibfnamefont {C.}~\bibnamefont {Gardiner}},\
  }\bibfield  {title} {\bibinfo {title} {Quantum theory of optical homodyne and
  heterodyne detection},\ }\href {https://doi.org/10.1080/09500348714550811}
  {\bibfield  {journal} {\bibinfo  {journal} {Journal of Modern Optics}\
  }\textbf {\bibinfo {volume} {34}},\ \bibinfo {pages} {881} (\bibinfo {year}
  {1987})}\BibitemShut {NoStop}%
\bibitem [{\citenamefont {Blow}\ \emph {et~al.}(1990)\citenamefont {Blow},
  \citenamefont {Loudon}, \citenamefont {Phoenix},\ and\ \citenamefont
  {Shepherd}}]{Blow1990}%
  \BibitemOpen
  \bibfield  {author} {\bibinfo {author} {\bibfnamefont {K.~J.}\ \bibnamefont
  {Blow}}, \bibinfo {author} {\bibfnamefont {R.}~\bibnamefont {Loudon}},
  \bibinfo {author} {\bibfnamefont {S.~J.~D.}\ \bibnamefont {Phoenix}},\ and\
  \bibinfo {author} {\bibfnamefont {T.~J.}\ \bibnamefont {Shepherd}},\
  }\bibfield  {title} {\bibinfo {title} {Continuum fields in quantum optics},\
  }\href {https://doi.org/10.1103/PhysRevA.42.4102} {\bibfield  {journal}
  {\bibinfo  {journal} {Physical Review A}\ }\textbf {\bibinfo {volume} {42}},\
  \bibinfo {pages} {4102} (\bibinfo {year} {1990})}\BibitemShut {NoStop}%
\bibitem [{\citenamefont {Bennink}\ and\ \citenamefont
  {Boyd}(2002)}]{Bennink2002}%
  \BibitemOpen
  \bibfield  {author} {\bibinfo {author} {\bibfnamefont {R.~S.}\ \bibnamefont
  {Bennink}}\ and\ \bibinfo {author} {\bibfnamefont {R.~W.}\ \bibnamefont
  {Boyd}},\ }\bibfield  {title} {\bibinfo {title} {{Improved measurement of
  multimode squeezed light via an eigenmode approach}},\ }\href
  {https://doi.org/10.1103/PhysRevA.66.053815} {\bibfield  {journal} {\bibinfo
  {journal} {Physical Review A}\ }\textbf {\bibinfo {volume} {66}},\ \bibinfo
  {pages} {5} (\bibinfo {year} {2002})}\BibitemShut {NoStop}%
\bibitem [{\citenamefont {Nakamura}(2021)}]{Nakamura2021}%
  \BibitemOpen
  \bibfield  {author} {\bibinfo {author} {\bibfnamefont {K.}~\bibnamefont
  {Nakamura}},\ }\bibfield  {title} {\bibinfo {title} {{Quantum noise and
  vacuum fluctuations in balanced homodyne detections through ideal multi-mode
  detectors}},\ }\href {https://doi.org/10.1093/ptep/ptab113} {\bibfield
  {journal} {\bibinfo  {journal} {Progress of Theoretical and Experimental
  Physics}\ }\textbf {\bibinfo {volume} {2021}},\ \bibinfo {pages} {103A01}
  (\bibinfo {year} {2021})}\BibitemShut {NoStop}%
\bibitem [{\citenamefont {Grosshans}\ and\ \citenamefont
  {Grangier}(2001)}]{Grosshans2001}%
  \BibitemOpen
  \bibfield  {author} {\bibinfo {author} {\bibfnamefont {F.}~\bibnamefont
  {Grosshans}}\ and\ \bibinfo {author} {\bibfnamefont {P.}~\bibnamefont
  {Grangier}},\ }\bibfield  {title} {\bibinfo {title} {{Effective quantum
  efficiency in the pulsed homodyne detection of a n-photon state}},\ }\href
  {https://doi.org/10.1007/s100530170243} {\bibfield  {journal} {\bibinfo
  {journal} {European Physical Journal D}\ }\textbf {\bibinfo {volume} {14}},\
  \bibinfo {pages} {119} (\bibinfo {year} {2001})}\BibitemShut {NoStop}%
\bibitem [{\citenamefont {Ou}(2017)}]{ou2017quantum}%
  \BibitemOpen
  \bibfield  {author} {\bibinfo {author} {\bibfnamefont {Z.}~\bibnamefont
  {Ou}},\ }\href {https://books.google.com/books?id=Ecw5DwAAQBAJ} {\emph
  {\bibinfo {title} {Quantum Optics For Experimentalists}}}\ (\bibinfo
  {publisher} {World Scientific Publishing Company},\ \bibinfo {year}
  {2017})\BibitemShut {NoStop}%
\bibitem [{\citenamefont {Chen}\ \emph {et~al.}(2023)\citenamefont {Chen},
  \citenamefont {Wang}, \citenamefont {Yu}, \citenamefont {Li},\ and\
  \citenamefont {Guo}}]{ChenWangYu2023}%
  \BibitemOpen
  \bibfield  {author} {\bibinfo {author} {\bibfnamefont {Z.}~\bibnamefont
  {Chen}}, \bibinfo {author} {\bibfnamefont {X.}~\bibnamefont {Wang}}, \bibinfo
  {author} {\bibfnamefont {S.}~\bibnamefont {Yu}}, \bibinfo {author}
  {\bibfnamefont {Z.}~\bibnamefont {Li}},\ and\ \bibinfo {author}
  {\bibfnamefont {H.}~\bibnamefont {Guo}},\ }\bibfield  {title} {\bibinfo
  {title} {Continuous-mode quantum key distribution with digital signal
  processing},\ }\href {https://doi.org/10.1038/s41534-023-00695-8} {\bibfield
  {journal} {\bibinfo  {journal} {npj Quantum Information}\ }\textbf {\bibinfo
  {volume} {9}},\ \bibinfo {pages} {28} (\bibinfo {year} {2023})}\BibitemShut
  {NoStop}%
\bibitem [{\citenamefont {Desch\^enes}\ and\ \citenamefont
  {Genest}(2013)}]{Deschenes2013}%
  \BibitemOpen
  \bibfield  {author} {\bibinfo {author} {\bibfnamefont {J.-D.}\ \bibnamefont
  {Desch\^enes}}\ and\ \bibinfo {author} {\bibfnamefont {J.}~\bibnamefont
  {Genest}},\ }\bibfield  {title} {\bibinfo {title} {Heterodyne beats between a
  continuous-wave laser and a frequency comb beyond the shot-noise limit of a
  single comb mode},\ }\href {https://doi.org/10.1103/PhysRevA.87.023802}
  {\bibfield  {journal} {\bibinfo  {journal} {Phys. Rev. A}\ }\textbf {\bibinfo
  {volume} {87}},\ \bibinfo {pages} {023802} (\bibinfo {year}
  {2013})}\BibitemShut {NoStop}%
\bibitem [{\citenamefont {Desch{\^{e}}nes}\ and\ \citenamefont
  {Genest}(2015)}]{Deschenes2015}%
  \BibitemOpen
  \bibfield  {author} {\bibinfo {author} {\bibfnamefont {J.-D.}\ \bibnamefont
  {Desch{\^{e}}nes}}\ and\ \bibinfo {author} {\bibfnamefont {J.}~\bibnamefont
  {Genest}},\ }\bibfield  {title} {\bibinfo {title} {{Chirped pulse heterodyne
  for optimal beat note detection between a frequency comb and a continuous
  wave laser}},\ }\href {https://doi.org/10.1364/OE.23.009295} {\bibfield
  {journal} {\bibinfo  {journal} {Opt. Express}\ }\textbf {\bibinfo {volume}
  {23}},\ \bibinfo {pages} {9295} (\bibinfo {year} {2015})}\BibitemShut
  {NoStop}%
\bibitem [{\citenamefont {Walsh}\ \emph {et~al.}(2023)\citenamefont {Walsh},
  \citenamefont {Guay},\ and\ \citenamefont {Genest}}]{Walsh2023}%
  \BibitemOpen
  \bibfield  {author} {\bibinfo {author} {\bibfnamefont {M.}~\bibnamefont
  {Walsh}}, \bibinfo {author} {\bibfnamefont {P.}~\bibnamefont {Guay}},\ and\
  \bibinfo {author} {\bibfnamefont {J.}~\bibnamefont {Genest}},\ }\bibfield
  {title} {\bibinfo {title} {Unlocking a lower shot noise limit in dual-comb
  interferometry},\ }\bibfield  {journal} {\bibinfo  {journal} {{APL}
  Photonics}\ }\textbf {\bibinfo {volume} {8}},\ \href
  {https://doi.org/10.1063/5.0153724} {10.1063/5.0153724} (\bibinfo {year}
  {2023})\BibitemShut {NoStop}%
\bibitem [{\citenamefont {Diddams}\ \emph {et~al.}(2020)\citenamefont
  {Diddams}, \citenamefont {Vahala},\ and\ \citenamefont {Udem}}]{Diddams2020}%
  \BibitemOpen
  \bibfield  {author} {\bibinfo {author} {\bibfnamefont {S.~A.}\ \bibnamefont
  {Diddams}}, \bibinfo {author} {\bibfnamefont {K.}~\bibnamefont {Vahala}},\
  and\ \bibinfo {author} {\bibfnamefont {T.}~\bibnamefont {Udem}},\ }\bibfield
  {title} {\bibinfo {title} {Optical frequency combs: Coherently uniting the
  electromagnetic spectrum},\ }\href {https://doi.org/10.1126/science.aay3676}
  {\bibfield  {journal} {\bibinfo  {journal} {Science}\ }\textbf {\bibinfo
  {volume} {369}},\ \bibinfo {pages} {eaay3676} (\bibinfo {year}
  {2020})}\BibitemShut {NoStop}%
\bibitem [{\citenamefont {Beloy}\ \emph {et~al.}(2021)\citenamefont {Beloy},
  \citenamefont {Bodine}, \citenamefont {Bothwell}, \citenamefont {Brewer},
  \citenamefont {Bromley}, \citenamefont {Chen}, \citenamefont {Desch{\^e}nes},
  \citenamefont {Diddams}, \citenamefont {Fasano}, \citenamefont {Fortier},
  \citenamefont {Hassan}, \citenamefont {Hume}, \citenamefont {Kedar},
  \citenamefont {Kennedy}, \citenamefont {Khader}, \citenamefont {Koepke},
  \citenamefont {Leibrandt}, \citenamefont {Leopardi}, \citenamefont {Ludlow},
  \citenamefont {McGrew}, \citenamefont {Milner}, \citenamefont {Newbury},
  \citenamefont {Nicolodi}, \citenamefont {Oelker}, \citenamefont {Parker},
  \citenamefont {Robinson}, \citenamefont {Romisch}, \citenamefont
  {Sch{\"a}ffer}, \citenamefont {Sherman}, \citenamefont {Sinclair},
  \citenamefont {Sonderhouse}, \citenamefont {Swann}, \citenamefont {Yao},
  \citenamefont {Ye}, \citenamefont {Zhang},\ and\ \citenamefont
  {Collaboration*}}]{Beloy2021}%
  \BibitemOpen
  \bibfield  {author} {\bibinfo {author} {\bibfnamefont {K.}~\bibnamefont
  {Beloy}}, \bibinfo {author} {\bibfnamefont {M.~I.}\ \bibnamefont {Bodine}},
  \bibinfo {author} {\bibfnamefont {T.}~\bibnamefont {Bothwell}}, \bibinfo
  {author} {\bibfnamefont {S.~M.}\ \bibnamefont {Brewer}}, \bibinfo {author}
  {\bibfnamefont {S.~L.}\ \bibnamefont {Bromley}}, \bibinfo {author}
  {\bibfnamefont {J.-S.}\ \bibnamefont {Chen}}, \bibinfo {author}
  {\bibfnamefont {J.-D.}\ \bibnamefont {Desch{\^e}nes}}, \bibinfo {author}
  {\bibfnamefont {S.~A.}\ \bibnamefont {Diddams}}, \bibinfo {author}
  {\bibfnamefont {R.~J.}\ \bibnamefont {Fasano}}, \bibinfo {author}
  {\bibfnamefont {T.~M.}\ \bibnamefont {Fortier}}, \bibinfo {author}
  {\bibfnamefont {Y.~S.}\ \bibnamefont {Hassan}}, \bibinfo {author}
  {\bibfnamefont {D.~B.}\ \bibnamefont {Hume}}, \bibinfo {author}
  {\bibfnamefont {D.}~\bibnamefont {Kedar}}, \bibinfo {author} {\bibfnamefont
  {C.~J.}\ \bibnamefont {Kennedy}}, \bibinfo {author} {\bibfnamefont
  {I.}~\bibnamefont {Khader}}, \bibinfo {author} {\bibfnamefont
  {A.}~\bibnamefont {Koepke}}, \bibinfo {author} {\bibfnamefont {D.~R.}\
  \bibnamefont {Leibrandt}}, \bibinfo {author} {\bibfnamefont {H.}~\bibnamefont
  {Leopardi}}, \bibinfo {author} {\bibfnamefont {A.~D.}\ \bibnamefont
  {Ludlow}}, \bibinfo {author} {\bibfnamefont {W.~F.}\ \bibnamefont {McGrew}},
  \bibinfo {author} {\bibfnamefont {W.~R.}\ \bibnamefont {Milner}}, \bibinfo
  {author} {\bibfnamefont {N.~R.}\ \bibnamefont {Newbury}}, \bibinfo {author}
  {\bibfnamefont {D.}~\bibnamefont {Nicolodi}}, \bibinfo {author}
  {\bibfnamefont {E.}~\bibnamefont {Oelker}}, \bibinfo {author} {\bibfnamefont
  {T.~E.}\ \bibnamefont {Parker}}, \bibinfo {author} {\bibfnamefont {J.~M.}\
  \bibnamefont {Robinson}}, \bibinfo {author} {\bibfnamefont {S.}~\bibnamefont
  {Romisch}}, \bibinfo {author} {\bibfnamefont {S.~A.}\ \bibnamefont
  {Sch{\"a}ffer}}, \bibinfo {author} {\bibfnamefont {J.~A.}\ \bibnamefont
  {Sherman}}, \bibinfo {author} {\bibfnamefont {L.~C.}\ \bibnamefont
  {Sinclair}}, \bibinfo {author} {\bibfnamefont {L.}~\bibnamefont
  {Sonderhouse}}, \bibinfo {author} {\bibfnamefont {W.~C.}\ \bibnamefont
  {Swann}}, \bibinfo {author} {\bibfnamefont {J.}~\bibnamefont {Yao}}, \bibinfo
  {author} {\bibfnamefont {J.}~\bibnamefont {Ye}}, \bibinfo {author}
  {\bibfnamefont {X.}~\bibnamefont {Zhang}},\ and\ \bibinfo {author}
  {\bibfnamefont {B.~A. C. O. N.~B.}\ \bibnamefont {Collaboration*}},\
  }\bibfield  {title} {\bibinfo {title} {Frequency ratio measurements at
  18-digit accuracy using an optical clock network},\ }\href
  {https://doi.org/10.1038/s41586-021-03253-4} {\bibfield  {journal} {\bibinfo
  {journal} {Nature}\ }\textbf {\bibinfo {volume} {591}},\ \bibinfo {pages}
  {564} (\bibinfo {year} {2021})}\BibitemShut {NoStop}%
\bibitem [{\citenamefont {Caldwell}\ \emph {et~al.}(2022)\citenamefont
  {Caldwell}, \citenamefont {Sinclair}, \citenamefont {Newbury},\ and\
  \citenamefont {Deschenes}}]{Caldwell2022}%
  \BibitemOpen
  \bibfield  {author} {\bibinfo {author} {\bibfnamefont {E.~D.}\ \bibnamefont
  {Caldwell}}, \bibinfo {author} {\bibfnamefont {L.~C.}\ \bibnamefont
  {Sinclair}}, \bibinfo {author} {\bibfnamefont {N.~R.}\ \bibnamefont
  {Newbury}},\ and\ \bibinfo {author} {\bibfnamefont {J.~D.}\ \bibnamefont
  {Deschenes}},\ }\bibfield  {title} {\bibinfo {title} {{The time-programmable
  frequency comb and its use in quantum-limited ranging}},\ }\href
  {https://doi.org/10.1038/s41586-022-05225-8} {\bibfield  {journal} {\bibinfo
  {journal} {Nature}\ }\textbf {\bibinfo {volume} {610}},\ \bibinfo {pages}
  {667} (\bibinfo {year} {2022})}\BibitemShut {NoStop}%
\bibitem [{\citenamefont {Shen}\ \emph {et~al.}(2022)\citenamefont {Shen},
  \citenamefont {Guan}, \citenamefont {Ren}, \citenamefont {Zeng},
  \citenamefont {Hou}, \citenamefont {Li}, \citenamefont {Cao}, \citenamefont
  {Han}, \citenamefont {Lian}, \citenamefont {Chen}, \citenamefont {Peng},
  \citenamefont {Wang}, \citenamefont {Zhu}, \citenamefont {Shi}, \citenamefont
  {Wang}, \citenamefont {Li}, \citenamefont {Liu}, \citenamefont {Pan},
  \citenamefont {Wang}, \citenamefont {Li}, \citenamefont {Wu}, \citenamefont
  {Zhang}, \citenamefont {Chen}, \citenamefont {Lu}, \citenamefont {Liao},
  \citenamefont {Yin}, \citenamefont {Jia}, \citenamefont {Peng}, \citenamefont
  {Jiang}, \citenamefont {Zhang},\ and\ \citenamefont {Pan}}]{Shen2022}%
  \BibitemOpen
  \bibfield  {author} {\bibinfo {author} {\bibfnamefont {Q.}~\bibnamefont
  {Shen}}, \bibinfo {author} {\bibfnamefont {J.~Y.}\ \bibnamefont {Guan}},
  \bibinfo {author} {\bibfnamefont {J.~G.}\ \bibnamefont {Ren}}, \bibinfo
  {author} {\bibfnamefont {T.}~\bibnamefont {Zeng}}, \bibinfo {author}
  {\bibfnamefont {L.}~\bibnamefont {Hou}}, \bibinfo {author} {\bibfnamefont
  {M.}~\bibnamefont {Li}}, \bibinfo {author} {\bibfnamefont {Y.}~\bibnamefont
  {Cao}}, \bibinfo {author} {\bibfnamefont {J.~J.}\ \bibnamefont {Han}},
  \bibinfo {author} {\bibfnamefont {M.~Z.}\ \bibnamefont {Lian}}, \bibinfo
  {author} {\bibfnamefont {Y.~W.}\ \bibnamefont {Chen}}, \bibinfo {author}
  {\bibfnamefont {X.~X.}\ \bibnamefont {Peng}}, \bibinfo {author}
  {\bibfnamefont {S.~M.}\ \bibnamefont {Wang}}, \bibinfo {author}
  {\bibfnamefont {D.~Y.}\ \bibnamefont {Zhu}}, \bibinfo {author} {\bibfnamefont
  {X.~P.}\ \bibnamefont {Shi}}, \bibinfo {author} {\bibfnamefont {Z.~G.}\
  \bibnamefont {Wang}}, \bibinfo {author} {\bibfnamefont {Y.}~\bibnamefont
  {Li}}, \bibinfo {author} {\bibfnamefont {W.~Y.}\ \bibnamefont {Liu}},
  \bibinfo {author} {\bibfnamefont {G.~S.}\ \bibnamefont {Pan}}, \bibinfo
  {author} {\bibfnamefont {Y.}~\bibnamefont {Wang}}, \bibinfo {author}
  {\bibfnamefont {Z.~H.}\ \bibnamefont {Li}}, \bibinfo {author} {\bibfnamefont
  {J.~C.}\ \bibnamefont {Wu}}, \bibinfo {author} {\bibfnamefont {Y.~Y.}\
  \bibnamefont {Zhang}}, \bibinfo {author} {\bibfnamefont {F.~X.}\ \bibnamefont
  {Chen}}, \bibinfo {author} {\bibfnamefont {C.~Y.}\ \bibnamefont {Lu}},
  \bibinfo {author} {\bibfnamefont {S.~K.}\ \bibnamefont {Liao}}, \bibinfo
  {author} {\bibfnamefont {J.}~\bibnamefont {Yin}}, \bibinfo {author}
  {\bibfnamefont {J.~J.}\ \bibnamefont {Jia}}, \bibinfo {author} {\bibfnamefont
  {C.~Z.}\ \bibnamefont {Peng}}, \bibinfo {author} {\bibfnamefont {H.~F.}\
  \bibnamefont {Jiang}}, \bibinfo {author} {\bibfnamefont {Q.}~\bibnamefont
  {Zhang}},\ and\ \bibinfo {author} {\bibfnamefont {J.~W.}\ \bibnamefont
  {Pan}},\ }\bibfield  {title} {\bibinfo {title} {Free-space dissemination of
  time and frequency with $10^{-19}$ instability over 113 km},\ }\href
  {https://doi.org/10.1038/s41586-022-05228-5} {\bibfield  {journal} {\bibinfo
  {journal} {Nature}\ }\textbf {\bibinfo {volume} {610}},\ \bibinfo {pages}
  {661} (\bibinfo {year} {2022})}\BibitemShut {NoStop}%
\bibitem [{\citenamefont {Coddington}\ \emph {et~al.}(2016)\citenamefont
  {Coddington}, \citenamefont {Newbury},\ and\ \citenamefont
  {Swann}}]{Coddington2016}%
  \BibitemOpen
  \bibfield  {author} {\bibinfo {author} {\bibfnamefont {I.}~\bibnamefont
  {Coddington}}, \bibinfo {author} {\bibfnamefont {N.}~\bibnamefont
  {Newbury}},\ and\ \bibinfo {author} {\bibfnamefont {W.}~\bibnamefont
  {Swann}},\ }\bibfield  {title} {\bibinfo {title} {{Dual-comb spectroscopy}},\
  }\href {https://doi.org/10.1364/optica.3.000414} {\bibfield  {journal}
  {\bibinfo  {journal} {Optica}\ }\textbf {\bibinfo {volume} {3}},\ \bibinfo
  {pages} {414} (\bibinfo {year} {2016})}\BibitemShut {NoStop}%
\bibitem [{\citenamefont {Picqu{\'e}}\ and\ \citenamefont
  {H{\"a}nsch}(2019)}]{Picque}%
  \BibitemOpen
  \bibfield  {author} {\bibinfo {author} {\bibfnamefont {N.}~\bibnamefont
  {Picqu{\'e}}}\ and\ \bibinfo {author} {\bibfnamefont {T.~W.}\ \bibnamefont
  {H{\"a}nsch}},\ }\bibfield  {title} {\bibinfo {title} {Frequency comb
  spectroscopy},\ }\href {https://doi.org/10.1038/s41566-018-0347-5} {\bibfield
   {journal} {\bibinfo  {journal} {Nature Photonics}\ }\textbf {\bibinfo
  {volume} {13}},\ \bibinfo {pages} {146} (\bibinfo {year} {2019})}\BibitemShut
  {NoStop}%
\bibitem [{\citenamefont {Yao}\ \emph {et~al.}(2016)\citenamefont {Yao},
  \citenamefont {Jiang}, \citenamefont {Yu}, \citenamefont {Bi},\ and\
  \citenamefont {Ma}}]{Yao2016}%
  \BibitemOpen
  \bibfield  {author} {\bibinfo {author} {\bibfnamefont {Y.}~\bibnamefont
  {Yao}}, \bibinfo {author} {\bibfnamefont {Y.}~\bibnamefont {Jiang}}, \bibinfo
  {author} {\bibfnamefont {H.}~\bibnamefont {Yu}}, \bibinfo {author}
  {\bibfnamefont {Z.}~\bibnamefont {Bi}},\ and\ \bibinfo {author}
  {\bibfnamefont {L.}~\bibnamefont {Ma}},\ }\bibfield  {title} {\bibinfo
  {title} {Optical frequency divider with division uncertainty at the
  $10^{-21}$ level},\ }\href {https://doi.org/10.1093/nsr/nww063} {\bibfield
  {journal} {\bibinfo  {journal} {National Science Review}\ }\textbf {\bibinfo
  {volume} {3}},\ \bibinfo {pages} {463} (\bibinfo {year} {2016})}\BibitemShut
  {NoStop}%
\bibitem [{\citenamefont {Xie}\ \emph {et~al.}(2017)\citenamefont {Xie},
  \citenamefont {Bouchand}, \citenamefont {Nicolodi}, \citenamefont {Giunta},
  \citenamefont {H{\"a}nsel}, \citenamefont {Lezius}, \citenamefont {Joshi},
  \citenamefont {Datta}, \citenamefont {Alexandre}, \citenamefont {Lours},
  \citenamefont {Tremblin}, \citenamefont {Santarelli}, \citenamefont
  {Holzwarth},\ and\ \citenamefont {Le~Coq}}]{Xie2016}%
  \BibitemOpen
  \bibfield  {author} {\bibinfo {author} {\bibfnamefont {X.}~\bibnamefont
  {Xie}}, \bibinfo {author} {\bibfnamefont {R.}~\bibnamefont {Bouchand}},
  \bibinfo {author} {\bibfnamefont {D.}~\bibnamefont {Nicolodi}}, \bibinfo
  {author} {\bibfnamefont {M.}~\bibnamefont {Giunta}}, \bibinfo {author}
  {\bibfnamefont {W.}~\bibnamefont {H{\"a}nsel}}, \bibinfo {author}
  {\bibfnamefont {M.}~\bibnamefont {Lezius}}, \bibinfo {author} {\bibfnamefont
  {A.}~\bibnamefont {Joshi}}, \bibinfo {author} {\bibfnamefont
  {S.}~\bibnamefont {Datta}}, \bibinfo {author} {\bibfnamefont
  {C.}~\bibnamefont {Alexandre}}, \bibinfo {author} {\bibfnamefont
  {M.}~\bibnamefont {Lours}}, \bibinfo {author} {\bibfnamefont {P.-A.}\
  \bibnamefont {Tremblin}}, \bibinfo {author} {\bibfnamefont {G.}~\bibnamefont
  {Santarelli}}, \bibinfo {author} {\bibfnamefont {R.}~\bibnamefont
  {Holzwarth}},\ and\ \bibinfo {author} {\bibfnamefont {Y.}~\bibnamefont
  {Le~Coq}},\ }\bibfield  {title} {\bibinfo {title} {Photonic microwave signals
  with zeptosecond-level absolute timing noise},\ }\href
  {https://doi.org/10.1038/nphoton.2016.215} {\bibfield  {journal} {\bibinfo
  {journal} {Nature Photonics}\ }\textbf {\bibinfo {volume} {11}},\ \bibinfo
  {pages} {44} (\bibinfo {year} {2017})}\BibitemShut {NoStop}%
\bibitem [{\citenamefont {Nakamura}\ \emph {et~al.}(2020)\citenamefont
  {Nakamura}, \citenamefont {Davila-Rodriguez}, \citenamefont {Leopardi},
  \citenamefont {Sherman}, \citenamefont {Fortier}, \citenamefont {Xie},
  \citenamefont {Campbell}, \citenamefont {McGrew}, \citenamefont {Zhang},
  \citenamefont {Hassan}, \citenamefont {Nicolodi}, \citenamefont {Beloy},
  \citenamefont {Ludlow}, \citenamefont {Diddams},\ and\ \citenamefont
  {Quinlan}}]{Nakamura2020b}%
  \BibitemOpen
  \bibfield  {author} {\bibinfo {author} {\bibfnamefont {T.}~\bibnamefont
  {Nakamura}}, \bibinfo {author} {\bibfnamefont {J.}~\bibnamefont
  {Davila-Rodriguez}}, \bibinfo {author} {\bibfnamefont {H.}~\bibnamefont
  {Leopardi}}, \bibinfo {author} {\bibfnamefont {J.~A.}\ \bibnamefont
  {Sherman}}, \bibinfo {author} {\bibfnamefont {T.~M.}\ \bibnamefont
  {Fortier}}, \bibinfo {author} {\bibfnamefont {X.}~\bibnamefont {Xie}},
  \bibinfo {author} {\bibfnamefont {J.~C.}\ \bibnamefont {Campbell}}, \bibinfo
  {author} {\bibfnamefont {W.~F.}\ \bibnamefont {McGrew}}, \bibinfo {author}
  {\bibfnamefont {X.}~\bibnamefont {Zhang}}, \bibinfo {author} {\bibfnamefont
  {Y.~S.}\ \bibnamefont {Hassan}}, \bibinfo {author} {\bibfnamefont
  {D.}~\bibnamefont {Nicolodi}}, \bibinfo {author} {\bibfnamefont
  {K.}~\bibnamefont {Beloy}}, \bibinfo {author} {\bibfnamefont {A.~D.}\
  \bibnamefont {Ludlow}}, \bibinfo {author} {\bibfnamefont {S.~A.}\
  \bibnamefont {Diddams}},\ and\ \bibinfo {author} {\bibfnamefont
  {F.}~\bibnamefont {Quinlan}},\ }\bibfield  {title} {\bibinfo {title}
  {Coherent optical clock down-conversion for microwave frequencies with
  $10^{-18}$ instability},\ }\href {https://doi.org/10.1126/science.abb2473}
  {\bibfield  {journal} {\bibinfo  {journal} {Science}\ }\textbf {\bibinfo
  {volume} {368}},\ \bibinfo {pages} {889} (\bibinfo {year}
  {2020})}\BibitemShut {NoStop}%
\bibitem [{\citenamefont {Reichert}\ \emph {et~al.}(1999)\citenamefont
  {Reichert}, \citenamefont {Holzwarth}, \citenamefont {Udem},\ and\
  \citenamefont {H{\"{a}}nsch}}]{REICHERT1999}%
  \BibitemOpen
  \bibfield  {author} {\bibinfo {author} {\bibfnamefont {J.}~\bibnamefont
  {Reichert}}, \bibinfo {author} {\bibfnamefont {R.}~\bibnamefont {Holzwarth}},
  \bibinfo {author} {\bibfnamefont {T.}~\bibnamefont {Udem}},\ and\ \bibinfo
  {author} {\bibfnamefont {T.~W.}\ \bibnamefont {H{\"{a}}nsch}},\ }\bibfield
  {title} {\bibinfo {title} {{Measuring the frequency of light with mode-locked
  lasers}},\ }\href {https://doi.org/10.1016/S0030-4018(99)00491-5} {\bibfield
  {journal} {\bibinfo  {journal} {Optics Communications}\ }\textbf {\bibinfo
  {volume} {172}},\ \bibinfo {pages} {59} (\bibinfo {year} {1999})}\BibitemShut
  {NoStop}%
\bibitem [{\citenamefont {Quinlan}\ \emph {et~al.}(2013)\citenamefont
  {Quinlan}, \citenamefont {Fortier}, \citenamefont {Jiang}, \citenamefont
  {Hati}, \citenamefont {Nelson}, \citenamefont {Fu}, \citenamefont
  {Campbell},\ and\ \citenamefont {Diddams}}]{Quinlan2013}%
  \BibitemOpen
  \bibfield  {author} {\bibinfo {author} {\bibfnamefont {F.}~\bibnamefont
  {Quinlan}}, \bibinfo {author} {\bibfnamefont {T.~M.}\ \bibnamefont
  {Fortier}}, \bibinfo {author} {\bibfnamefont {H.}~\bibnamefont {Jiang}},
  \bibinfo {author} {\bibfnamefont {A.}~\bibnamefont {Hati}}, \bibinfo {author}
  {\bibfnamefont {C.}~\bibnamefont {Nelson}}, \bibinfo {author} {\bibfnamefont
  {Y.}~\bibnamefont {Fu}}, \bibinfo {author} {\bibfnamefont {J.~C.}\
  \bibnamefont {Campbell}},\ and\ \bibinfo {author} {\bibfnamefont {S.~A.}\
  \bibnamefont {Diddams}},\ }\bibfield  {title} {\bibinfo {title} {Exploiting
  shot noise correlations in the photodetection of ultrashort optical pulse
  trains},\ }\href {https://doi.org/10.1038/nphoton.2013.33} {\bibfield
  {journal} {\bibinfo  {journal} {Nature Photonics}\ }\textbf {\bibinfo
  {volume} {7}},\ \bibinfo {pages} {290} (\bibinfo {year} {2013})}\BibitemShut
  {NoStop}%
\bibitem [{\citenamefont {Pinel}\ \emph {et~al.}(2012)\citenamefont {Pinel},
  \citenamefont {Jian}, \citenamefont {de~Ara\'ujo}, \citenamefont {Feng},
  \citenamefont {Chalopin}, \citenamefont {Fabre},\ and\ \citenamefont
  {Treps}}]{PinelJianTreps2012}%
  \BibitemOpen
  \bibfield  {author} {\bibinfo {author} {\bibfnamefont {O.}~\bibnamefont
  {Pinel}}, \bibinfo {author} {\bibfnamefont {P.}~\bibnamefont {Jian}},
  \bibinfo {author} {\bibfnamefont {R.~M.}\ \bibnamefont {de~Ara\'ujo}},
  \bibinfo {author} {\bibfnamefont {J.}~\bibnamefont {Feng}}, \bibinfo {author}
  {\bibfnamefont {B.}~\bibnamefont {Chalopin}}, \bibinfo {author}
  {\bibfnamefont {C.}~\bibnamefont {Fabre}},\ and\ \bibinfo {author}
  {\bibfnamefont {N.}~\bibnamefont {Treps}},\ }\bibfield  {title} {\bibinfo
  {title} {Generation and characterization of multimode quantum frequency
  combs},\ }\href {https://doi.org/10.1103/PhysRevLett.108.083601} {\bibfield
  {journal} {\bibinfo  {journal} {Phys. Rev. Lett.}\ }\textbf {\bibinfo
  {volume} {108}},\ \bibinfo {pages} {083601} (\bibinfo {year}
  {2012})}\BibitemShut {NoStop}%
\bibitem [{\citenamefont {Kues}\ \emph {et~al.}(2019)\citenamefont {Kues},
  \citenamefont {Reimer}, \citenamefont {Lukens}, \citenamefont {Munro},
  \citenamefont {Weiner}, \citenamefont {Moss},\ and\ \citenamefont
  {Morandotti}}]{KuesMorandotti2019}%
  \BibitemOpen
  \bibfield  {author} {\bibinfo {author} {\bibfnamefont {M.}~\bibnamefont
  {Kues}}, \bibinfo {author} {\bibfnamefont {C.}~\bibnamefont {Reimer}},
  \bibinfo {author} {\bibfnamefont {J.~M.}\ \bibnamefont {Lukens}}, \bibinfo
  {author} {\bibfnamefont {W.~J.}\ \bibnamefont {Munro}}, \bibinfo {author}
  {\bibfnamefont {A.~M.}\ \bibnamefont {Weiner}}, \bibinfo {author}
  {\bibfnamefont {D.~J.}\ \bibnamefont {Moss}},\ and\ \bibinfo {author}
  {\bibfnamefont {R.}~\bibnamefont {Morandotti}},\ }\bibfield  {title}
  {\bibinfo {title} {Quantum optical microcombs},\ }\href
  {https://doi.org/10.1038/s41566-019-0363-0} {\bibfield  {journal} {\bibinfo
  {journal} {Nature Photonics}\ }\textbf {\bibinfo {volume} {13}},\ \bibinfo
  {pages} {170} (\bibinfo {year} {2019})}\BibitemShut {NoStop}%
\bibitem [{\citenamefont {Fabre}\ \emph {et~al.}(2020)\citenamefont {Fabre},
  \citenamefont {Maltese}, \citenamefont {Appas}, \citenamefont {Felicetti},
  \citenamefont {Ketterer}, \citenamefont {Keller}, \citenamefont {Coudreau},
  \citenamefont {Baboux}, \citenamefont {Amanti}, \citenamefont {Ducci},\ and\
  \citenamefont {Milman}}]{FabreMilman2020}%
  \BibitemOpen
  \bibfield  {author} {\bibinfo {author} {\bibfnamefont {N.}~\bibnamefont
  {Fabre}}, \bibinfo {author} {\bibfnamefont {G.}~\bibnamefont {Maltese}},
  \bibinfo {author} {\bibfnamefont {F.}~\bibnamefont {Appas}}, \bibinfo
  {author} {\bibfnamefont {S.}~\bibnamefont {Felicetti}}, \bibinfo {author}
  {\bibfnamefont {A.}~\bibnamefont {Ketterer}}, \bibinfo {author}
  {\bibfnamefont {A.}~\bibnamefont {Keller}}, \bibinfo {author} {\bibfnamefont
  {T.}~\bibnamefont {Coudreau}}, \bibinfo {author} {\bibfnamefont
  {F.}~\bibnamefont {Baboux}}, \bibinfo {author} {\bibfnamefont {M.~I.}\
  \bibnamefont {Amanti}}, \bibinfo {author} {\bibfnamefont {S.}~\bibnamefont
  {Ducci}},\ and\ \bibinfo {author} {\bibfnamefont {P.}~\bibnamefont
  {Milman}},\ }\bibfield  {title} {\bibinfo {title} {Generation of a
  time-frequency grid state with integrated biphoton frequency combs},\ }\href
  {https://doi.org/10.1103/PhysRevA.102.012607} {\bibfield  {journal} {\bibinfo
   {journal} {Phys. Rev. A}\ }\textbf {\bibinfo {volume} {102}},\ \bibinfo
  {pages} {012607} (\bibinfo {year} {2020})}\BibitemShut {NoStop}%
\bibitem [{\citenamefont {Maltese}\ \emph {et~al.}(2020)\citenamefont
  {Maltese}, \citenamefont {Amanti}, \citenamefont {Appas}, \citenamefont
  {Sinnl}, \citenamefont {Lema{\^\i}tre}, \citenamefont {Milman}, \citenamefont
  {Baboux},\ and\ \citenamefont {Ducci}}]{MalteseDucci2020}%
  \BibitemOpen
  \bibfield  {author} {\bibinfo {author} {\bibfnamefont {G.}~\bibnamefont
  {Maltese}}, \bibinfo {author} {\bibfnamefont {M.~I.}\ \bibnamefont {Amanti}},
  \bibinfo {author} {\bibfnamefont {F.}~\bibnamefont {Appas}}, \bibinfo
  {author} {\bibfnamefont {G.}~\bibnamefont {Sinnl}}, \bibinfo {author}
  {\bibfnamefont {A.}~\bibnamefont {Lema{\^\i}tre}}, \bibinfo {author}
  {\bibfnamefont {P.}~\bibnamefont {Milman}}, \bibinfo {author} {\bibfnamefont
  {F.}~\bibnamefont {Baboux}},\ and\ \bibinfo {author} {\bibfnamefont
  {S.}~\bibnamefont {Ducci}},\ }\bibfield  {title} {\bibinfo {title}
  {Generation and symmetry control of quantum frequency combs},\ }\href
  {https://doi.org/10.1038/s41534-019-0237-9} {\bibfield  {journal} {\bibinfo
  {journal} {npj Quantum Information}\ }\textbf {\bibinfo {volume} {6}},\
  \bibinfo {pages} {13} (\bibinfo {year} {2020})}\BibitemShut {NoStop}%
\bibitem [{\citenamefont {Yang}\ \emph {et~al.}(2021)\citenamefont {Yang},
  \citenamefont {Jahanbozorgi}, \citenamefont {Jeong}, \citenamefont {Sun},
  \citenamefont {Pfister}, \citenamefont {Lee},\ and\ \citenamefont
  {Yi}}]{YangYi2021}%
  \BibitemOpen
  \bibfield  {author} {\bibinfo {author} {\bibfnamefont {Z.}~\bibnamefont
  {Yang}}, \bibinfo {author} {\bibfnamefont {M.}~\bibnamefont {Jahanbozorgi}},
  \bibinfo {author} {\bibfnamefont {D.}~\bibnamefont {Jeong}}, \bibinfo
  {author} {\bibfnamefont {S.}~\bibnamefont {Sun}}, \bibinfo {author}
  {\bibfnamefont {O.}~\bibnamefont {Pfister}}, \bibinfo {author} {\bibfnamefont
  {H.}~\bibnamefont {Lee}},\ and\ \bibinfo {author} {\bibfnamefont
  {X.}~\bibnamefont {Yi}},\ }\bibfield  {title} {\bibinfo {title} {A squeezed
  quantum microcomb on a chip},\ }\href
  {https://doi.org/10.1038/s41467-021-25054-z} {\bibfield  {journal} {\bibinfo
  {journal} {Nature Communications}\ }\textbf {\bibinfo {volume} {12}},\
  \bibinfo {pages} {4781} (\bibinfo {year} {2021})}\BibitemShut {NoStop}%
\bibitem [{\citenamefont {Cai}\ \emph {et~al.}(2021)\citenamefont {Cai},
  \citenamefont {Roslund}, \citenamefont {Thiel}, \citenamefont {Fabre},\ and\
  \citenamefont {Treps}}]{CaiRoslundTreps2021}%
  \BibitemOpen
  \bibfield  {author} {\bibinfo {author} {\bibfnamefont {Y.}~\bibnamefont
  {Cai}}, \bibinfo {author} {\bibfnamefont {J.}~\bibnamefont {Roslund}},
  \bibinfo {author} {\bibfnamefont {V.}~\bibnamefont {Thiel}}, \bibinfo
  {author} {\bibfnamefont {C.}~\bibnamefont {Fabre}},\ and\ \bibinfo {author}
  {\bibfnamefont {N.}~\bibnamefont {Treps}},\ }\bibfield  {title} {\bibinfo
  {title} {Quantum enhanced measurement of an optical frequency comb},\ }\href
  {https://doi.org/10.1038/s41534-021-00419-w} {\bibfield  {journal} {\bibinfo
  {journal} {npj Quantum Information}\ }\textbf {\bibinfo {volume} {7}},\
  \bibinfo {pages} {82} (\bibinfo {year} {2021})}\BibitemShut {NoStop}%
\bibitem [{\citenamefont {Belsley}(2023)}]{Belsley2023}%
  \BibitemOpen
  \bibfield  {author} {\bibinfo {author} {\bibfnamefont {A.}~\bibnamefont
  {Belsley}},\ }\bibfield  {title} {\bibinfo {title} {Quantum-enhanced
  absorption spectroscopy with bright squeezed frequency combs},\ }\href
  {https://doi.org/10.1103/PhysRevLett.130.133602} {\bibfield  {journal}
  {\bibinfo  {journal} {Phys. Rev. Lett.}\ }\textbf {\bibinfo {volume} {130}},\
  \bibinfo {pages} {133602} (\bibinfo {year} {2023})}\BibitemShut {NoStop}%
\bibitem [{\citenamefont {Guidry}\ \emph {et~al.}(2023)\citenamefont {Guidry},
  \citenamefont {Lukin}, \citenamefont {Yang},\ and\ \citenamefont
  {Vu\v{c}kovi\'{c}}}]{GuidryVuckovic2023}%
  \BibitemOpen
  \bibfield  {author} {\bibinfo {author} {\bibfnamefont {M.~A.}\ \bibnamefont
  {Guidry}}, \bibinfo {author} {\bibfnamefont {D.~M.}\ \bibnamefont {Lukin}},
  \bibinfo {author} {\bibfnamefont {K.~Y.}\ \bibnamefont {Yang}},\ and\
  \bibinfo {author} {\bibfnamefont {J.}~\bibnamefont {Vu\v{c}kovi\'{c}}},\
  }\bibfield  {title} {\bibinfo {title} {Multimode squeezing in soliton crystal
  microcombs},\ }\href {https://doi.org/10.1364/OPTICA.485996} {\bibfield
  {journal} {\bibinfo  {journal} {Optica}\ }\textbf {\bibinfo {volume} {10}},\
  \bibinfo {pages} {694} (\bibinfo {year} {2023})}\BibitemShut {NoStop}%
\bibitem [{\citenamefont {Shi}\ \emph {et~al.}(2023)\citenamefont {Shi},
  \citenamefont {Chen}, \citenamefont {Fraser}, \citenamefont {Yu},
  \citenamefont {Zhang},\ and\ \citenamefont {Zhuang}}]{ShiChenZhuang2023}%
  \BibitemOpen
  \bibfield  {author} {\bibinfo {author} {\bibfnamefont {H.}~\bibnamefont
  {Shi}}, \bibinfo {author} {\bibfnamefont {Z.}~\bibnamefont {Chen}}, \bibinfo
  {author} {\bibfnamefont {S.~E.}\ \bibnamefont {Fraser}}, \bibinfo {author}
  {\bibfnamefont {M.}~\bibnamefont {Yu}}, \bibinfo {author} {\bibfnamefont
  {Z.}~\bibnamefont {Zhang}},\ and\ \bibinfo {author} {\bibfnamefont
  {Q.}~\bibnamefont {Zhuang}},\ }\bibfield  {title} {\bibinfo {title}
  {Entanglement-enhanced dual-comb spectroscopy},\ }\bibfield  {journal}
  {\bibinfo  {journal} {{arXiv}}\ }\href
  {https://doi.org/10.48550/arXiv.2304.01516} {10.48550/arXiv.2304.01516}
  (\bibinfo {year} {2023})\BibitemShut {NoStop}%
\bibitem [{\citenamefont {Loudon}(2000)}]{loudon2000quantum}%
  \BibitemOpen
  \bibfield  {author} {\bibinfo {author} {\bibfnamefont {R.}~\bibnamefont
  {Loudon}},\ }\href@noop {} {\emph {\bibinfo {title} {The quantum theory of
  light}}},\ \bibinfo {edition} {3rd}\ ed.\ (\bibinfo  {publisher} {OUP
  Oxford},\ \bibinfo {year} {2000})\BibitemShut {NoStop}%
\bibitem [{\citenamefont {Combes}\ and\ \citenamefont
  {Lund}(2022)}]{Combes2022}%
  \BibitemOpen
  \bibfield  {author} {\bibinfo {author} {\bibfnamefont {J.}~\bibnamefont
  {Combes}}\ and\ \bibinfo {author} {\bibfnamefont {A.~P.}\ \bibnamefont
  {Lund}},\ }\bibfield  {title} {\bibinfo {title} {{Homodyne measurement with a
  Schr\"odinger cat state as a local oscillator}},\ }\href
  {https://doi.org/10.1103/PhysRevA.106.063706} {\bibfield  {journal} {\bibinfo
   {journal} {Physical Review A}\ }\textbf {\bibinfo {volume} {106}},\ \bibinfo
  {pages} {063706} (\bibinfo {year} {2022})}\BibitemShut {NoStop}%
\bibitem [{\citenamefont {Avagyan}(2023)}]{avagyan_quantum_2023}%
  \BibitemOpen
  \bibfield  {author} {\bibinfo {author} {\bibfnamefont {A.}~\bibnamefont
  {Avagyan}},\ }\href {https://doi.org/10.48550/arXiv.2305.19397} {\bibinfo
  {title} {Quantum {State} {Characterization} {Using} {Measurement}
  {Configurations} {Inspired} by {Homodyne} {Detection}}} (\bibinfo {year}
  {2023}),\ \bibinfo {note} {arXiv:2305.19397 [quant-ph]}\BibitemShut {NoStop}%
\bibitem [{\citenamefont {Oh}(2014)}]{oh_maximum_2014}%
  \BibitemOpen
  \bibfield  {author} {\bibinfo {author} {\bibfnamefont {C.~H.}\ \bibnamefont
  {Oh}},\ }\bibfield  {title} {\bibinfo {title} {A maximum likelihood
  estimation method for a mixture of shifted binomial distributions},\ }\href
  {https://www.dbpia.co.kr/Journal/articleDetail?nodeId=NODE07244978}
  {\bibfield  {journal} {\bibinfo  {journal} {Korean Journal of Data
  Information Science}\ }\textbf {\bibinfo {volume} {25}},\ \bibinfo {pages}
  {255} (\bibinfo {year} {2014})}\BibitemShut {NoStop}%
\bibitem [{\citenamefont {Vogel}(1995)}]{Vogel1995}%
  \BibitemOpen
  \bibfield  {author} {\bibinfo {author} {\bibfnamefont {W.}~\bibnamefont
  {Vogel}},\ }\bibfield  {title} {\bibinfo {title} {Homodyne correlation
  measurements with weak local oscillators},\ }\href
  {https://doi.org/10.1103/PhysRevA.51.4160} {\bibfield  {journal} {\bibinfo
  {journal} {Phys. Rev. A}\ }\textbf {\bibinfo {volume} {51}},\ \bibinfo
  {pages} {4160} (\bibinfo {year} {1995})}\BibitemShut {NoStop}%
\bibitem [{\citenamefont {Thekkadath}\ \emph {et~al.}(2020)\citenamefont
  {Thekkadath}, \citenamefont {Phillips}, \citenamefont {Bulmer}, \citenamefont
  {Clements}, \citenamefont {Eckstein}, \citenamefont {Bell}, \citenamefont
  {Lugani}, \citenamefont {Wolterink}, \citenamefont {Lita}, \citenamefont
  {Nam}, \citenamefont {Gerrits}, \citenamefont {Wade},\ and\ \citenamefont
  {Walmsley}}]{ThekkadathWalmsley2020}%
  \BibitemOpen
  \bibfield  {author} {\bibinfo {author} {\bibfnamefont {G.~S.}\ \bibnamefont
  {Thekkadath}}, \bibinfo {author} {\bibfnamefont {D.~S.}\ \bibnamefont
  {Phillips}}, \bibinfo {author} {\bibfnamefont {J.~F.~F.}\ \bibnamefont
  {Bulmer}}, \bibinfo {author} {\bibfnamefont {W.~R.}\ \bibnamefont
  {Clements}}, \bibinfo {author} {\bibfnamefont {A.}~\bibnamefont {Eckstein}},
  \bibinfo {author} {\bibfnamefont {B.~A.}\ \bibnamefont {Bell}}, \bibinfo
  {author} {\bibfnamefont {J.}~\bibnamefont {Lugani}}, \bibinfo {author}
  {\bibfnamefont {T.~A.~W.}\ \bibnamefont {Wolterink}}, \bibinfo {author}
  {\bibfnamefont {A.}~\bibnamefont {Lita}}, \bibinfo {author} {\bibfnamefont
  {S.~W.}\ \bibnamefont {Nam}}, \bibinfo {author} {\bibfnamefont
  {T.}~\bibnamefont {Gerrits}}, \bibinfo {author} {\bibfnamefont {C.~G.}\
  \bibnamefont {Wade}},\ and\ \bibinfo {author} {\bibfnamefont {I.~A.}\
  \bibnamefont {Walmsley}},\ }\bibfield  {title} {\bibinfo {title} {Tuning
  between photon-number and quadrature measurements with weak-field homodyne
  detection},\ }\href {https://doi.org/10.1103/PhysRevA.101.031801} {\bibfield
  {journal} {\bibinfo  {journal} {Phys. Rev. A}\ }\textbf {\bibinfo {volume}
  {101}},\ \bibinfo {pages} {031801} (\bibinfo {year} {2020})}\BibitemShut
  {NoStop}%
\bibitem [{\citenamefont {Olivares}\ \emph {et~al.}(2020)\citenamefont
  {Olivares}, \citenamefont {Allevi},\ and\ \citenamefont
  {Bondani}}]{OlivaresBondani2020}%
  \BibitemOpen
  \bibfield  {author} {\bibinfo {author} {\bibfnamefont {S.}~\bibnamefont
  {Olivares}}, \bibinfo {author} {\bibfnamefont {A.}~\bibnamefont {Allevi}},\
  and\ \bibinfo {author} {\bibfnamefont {M.}~\bibnamefont {Bondani}},\
  }\bibfield  {title} {\bibinfo {title} {On the role of the local oscillator
  intensity in optical homodyne-like tomography},\ }\href
  {https://doi.org/https://doi.org/10.1016/j.physleta.2020.126354} {\bibfield
  {journal} {\bibinfo  {journal} {Physics Letters A}\ }\textbf {\bibinfo
  {volume} {384}},\ \bibinfo {pages} {126354} (\bibinfo {year}
  {2020})}\BibitemShut {NoStop}%
\bibitem [{\citenamefont {Hubenschmid}\ \emph {et~al.}(2022)\citenamefont
  {Hubenschmid}, \citenamefont {Guedes},\ and\ \citenamefont
  {Burkard}}]{HubenschmidBurkard2022}%
  \BibitemOpen
  \bibfield  {author} {\bibinfo {author} {\bibfnamefont {E.}~\bibnamefont
  {Hubenschmid}}, \bibinfo {author} {\bibfnamefont {T.~L.~M.}\ \bibnamefont
  {Guedes}},\ and\ \bibinfo {author} {\bibfnamefont {G.}~\bibnamefont
  {Burkard}},\ }\bibfield  {title} {\bibinfo {title} {Complete positive
  operator-valued measure description of multichannel quantum electro-optic
  sampling with monochromatic field modes},\ }\href
  {https://doi.org/10.1103/PhysRevA.106.043713} {\bibfield  {journal} {\bibinfo
   {journal} {Phys. Rev. A}\ }\textbf {\bibinfo {volume} {106}},\ \bibinfo
  {pages} {043713} (\bibinfo {year} {2022})}\BibitemShut {NoStop}%
\end{thebibliography}%

\vspace{10pt}

\appendix
\newpage

% ========================================================
\section{Mode Decomposition of Beamsplitter}\label{app:Beamsplitter}
% ========================================================
We show the the beamsplitter acts the same for every mode, under the assumption that the transmission and reflection coefficients of the beamsplitter are constants over the relevant frequencies of the signal and LO mode. From our assumption we have the the action of the beamsplitter in every mode is described by the following input output relations,
\begin{subequations}\label{eqn:beamsplitter_action}
\begin{align}
         U_{\text{BS}}^\dagger(\xi)\hat{A}(\xi) U_{\text{BS}}(\xi) &=\frac{\hat{A}(\xi) + \hat{B}(\xi)}{\sqrt{2}} \\
         U_{\text{BS}}^\dagger(\xi)\hat{B}(\xi) U_{\text{BS}}(\xi) &=\frac{\hat{A}(\xi) - \hat{B}(\xi)}{\sqrt{2}}\, .
\end{align}
\end{subequations}
We want to show that this implies a tensor product structure to the beamsplitter unitary, i.e we can split the unitary into a beamsplitter for each mode independently. We can demonstrate this by considering decomposing a single mode into a combination of modes and seeing how the unitary must act. Consider a annihilation operator in some mode $\hat{A}(\xi)$. Further allow $\xi$ to be decomposed as $\xi = c_1\xi_1 +c_2\xi_2+\dots$.
When we conjugate $\hat{A}(\xi)$ by the beamsplitter we have,
\begin{equation}
    \begin{aligned}
        U_{\text{BS}}^\dagger\hat{A}(\xi) U_{\text{BS}} &=  U_{\text{BS}}^\dagger\left(\sum_n c_n\hat{A}(\xi_n)\right) U_{\text{BS}}\, .
    \end{aligned}
\end{equation}
but from \cref{eqn:beamsplitter_action} we know this must produce a sum of operators in the $\xi$ mode. 
\begin{equation}
    U_{\text{BS}}^\dagger\left(\sum_n c_n\hat{A}(\xi_n)\right) U_{\text{BS}} = \frac{1}{\sqrt{2}} \sum_n c_n\big( \hat{A}(\xi_n)+\hat{B}(\xi_n) \big ) \,.
\end{equation}
In order for this to hold we need the following to be true,
\begin{equation}
    U_{\text{BS}}^{ {\dagger} } \hat{A}(\xi_n)U_{\text{BS}} = \frac{\hat{A}(\xi_n)+\hat{B}(\xi_n)}{\sqrt{2}}.
\end{equation}
This equation allows us to write $U_{\text{BS}}(\xi)$ by its action on every mode. Meaning we can decompose the beamsplitter unitary into modes as 
\begin{equation}
    U_{\text{BS}} = U_{\text{BS}}(\xi_1)\otimes U_{\text{BS}}(\xi_2)\otimes \dots,
\end{equation}
as desired.

% ========================================================
\section{Deriving the Single Mode POVMs from Kraus Operators}\label{app:POVM}
% ========================================================

\subsection{Perpendicular mode}\label{app:POVM_perp}
First we address the perpendicular mode, which is significantly easier. We start with equation \cref{eqn:perp_Int}
\begin{equation}
     M_{r,s}(\xi_\perp) = \<r_{\xi_\perp}|\<s_{\xi_\perp}|U_{\text{BS}}|\vac\>\, ,
\end{equation}
which represents $r$ and $s$ clicks in the perpendicular mode. Using the definition of photon number states from \cref{eqn:fock} to expand the bras into operators acting of vacuum
\begin{equation}
    M_{r,s}(\xi_\perp) = \<\vac|\<\vac|\frac{\hat{A}^r\hat{B}^s}{\sqrt{r!s!}}U_{\text{BS}}|\vac\>.
\end{equation}
We have omitted the mode label since every operator in this equation acts on the perpendicular mode. We now insert identity, i.e. $U_{\text{BS}}^\dagger U_{\text{BS}} = I$,  after every operator so we can apply the input-output relations of the beamsplitter, $U_{\text{BS}}^\dagger \hat{A}U_{\text{BS}} =(\hat{A} + \hat{B})/\sqrt{2} $ and  $U_{\text{BS}}^\dagger \hat{B}U_{\text{BS}} =(\hat{A} - \hat{B})/\sqrt{2} $. Doing this and applying the remaining beamsplitter unitary to the vacuum modes on the left we get the following equation
\begin{equation}\label{eqn:Beamsplitter}
      M_{r,s}(\xi_\perp) = \<\vac|\<\vac|\frac{\left(\hat{A}+\hat{B}\right)^r\left(\hat{A}-\hat{B}\right)^s}{\sqrt{r!s!}2^{(r+s)/2}}|\vac\>
\end{equation}
where the $\hat{B}$ operator acts on the Hilbert space of the LO, which in this case is in vacuum. We can apply this vacuum state effectively replacing each $\hat{B}$ operator with a $0$ and get,
\begin{align}
     M_{r,s}(\xi_\perp) &= \<\vac|\frac{\hat{A}^{r+s}}{\sqrt{r!s!}2^{(r+s)/2}}\nonumber\\
     &=\<r+s|\frac{\sqrt{(r+s)!}}{\sqrt{r!s!}2^{(r+s)/2}} \nonumber\\
     &=\<r+s|\frac{1}{2^{(r+s)/2}}\sqrt{{r+s \choose r}}
\end{align}

Now we move to the POVMs $E_{r,s} = M_{r,s}^\dagger M_{r,s}$
\begin{equation}
    E_{r,s} = \frac{|r+s\>\<r+s|}{2^{(r+s)}}{r+s \choose r} \, ,
\end{equation}
and change to sum and difference variables as follows,
\begin{equation}
    E_{x,w} = \frac{|w\>\<w|}{2^{w}}{w \choose \frac{\tilde{x}}{2}+\frac{w}{2}}.
\end{equation}
This is a shifted binomial distribution in $x$ with mean 0 in the difference variable and variance $w$. While we could further approximate this distribution in the limit where $w$ is large we choose not to for two reasons. Firstly, we want the mode-matched limit $\gamma = 1$ to appear naturally from our results and in that limit $w$ in the perpendicular mode goes to zero. Second, the binomial distribution has some nice properties particularly when coupled with the Poisson distribution of a coherent input state that make the marginalization integrals analytically solvable. For these reasons we will scale the difference variable, but leave it discrete at the POVM level. Depending on the input state the difference variable can be made continuous in a variety of ways, most commonly by approximating the binomial distribution as normal. The resulting marginalization over w will be difficult under this approximation but can easliy be solved numberically. % this as an exercise for the reader to determine the meaning of this word
After appropriately scaling variables we get the following POVM:
\begin{equation}
    E_{x,w}dx = dx|w\>\<w|\text{Bin}\Big(\frac{|\beta|x}{\sqrt{2}}\Big|w,\frac{1}{2},0\Big)
\end{equation}
where $\text{Bin}\big(x\big|n,p,\mu\big)$ is a  binomial distribution characterised by $n$ and $p$ and shifted so that it has mean $\mu$
\begin{equation}
    \text{Bin}\big(x\big|n,p,\mu\big) = {n \choose x+\frac{n}{2}-\mu}p^{n/2+x-\mu}(1-p)^{n/2-x+\mu} .
\end{equation}

\subsection{LO mode}\label{app:POVM_LO}

The calculation in the LO mode is more involved, but has been described in detail in \cite{Combes2022}.  We start with \cref{eqn:LO_Quad},
\begin{equation}
     M_{p,q}(\xi_\lo) = \<p_{\xi_\lo}|\<q_{\xi_\lo}|U_{\text{BS}}|\beta_{\xi_\lo}\> \, .
\end{equation}
We apply the same steps as before up to \cref{eqn:Beamsplitter}, the only difference being the state of the LO is no longer vacuum. Doing this yields
\begin{equation}
    M_{p,q} = \<\vac|\<\vac|\frac{\left(\hat{A}+\hat{B}\right)^p\left(\hat{A}-\hat{B}\right)^q}{\sqrt{p!q!}2^{(p+q)/2}}|\beta\> \, ,
\end{equation}
where the mode designations are omitted because every operator and state is in the LO mode. After acting operators on the LO coherent state we arrive at
\begin{equation}
     M_{p,q} = \<\vac|\frac{\left(\hat{A}+\beta\right)^p\left(\hat{A}-\beta\right)^q}{\sqrt{p!q!}2^{(p+q)/2}}e^{-|\beta|^2/2}.
\end{equation}
Now this operator acts only on the signal Hilbert space.

From here we apply a series of algebraic manipulations to arrange the operator into a form where we can apply the large local oscillator assumption. When we do this we will assume without loss of generality that $p\ge q$, but the calculation goes much the same with the opposite assumption. After manipulation we get 
\begin{align}
      M_{p,q} =\<\vac| &(-1)^q\left(1+\frac{\hat{A}}{\beta}\right)^{p-q}\left(1-\frac{\hat{A}^2}{\beta^2}\right)^q \times \nonumber\\
      &\frac{e^{-|\beta|^2/4}}{\sqrt{p!}}\left(\frac{\beta}{\sqrt{2}}\right)^p\frac{e^{-|\beta|^2/4}}{\sqrt{q!}}\left(\frac{\beta}{\sqrt{2}}\right)^q
\end{align}
We have arranged these terms so that we see the appearance of two Poisson distributions in $p$ and $q$ as well as two terms that can be expanded in the large $\beta$ limit into exponentials. From here we again move partially to the sum and difference variables of \cref{eqn:sumdiff} and also replace $m$ with its mean $|\beta|^2/2$. Applying the large LO limit allows us to approximate these Poisson distributions as normal as well. Doing this will move us from discrete variables $p$ and $q$ to continuous variables $p'$ and $q'$. Applying all of these leads us to
\begin{align}
       M_{p',q'} \approx& \<\vac|(-1)^{q'}e^{\sqrt{2}e^{-i\theta}x\hat{A}}e^{-e^{-2i\theta}\hat{A}^2/2} \times\nonumber\\
       &\left(\frac{e^{-(p'-|\beta|^2/2)^2/(2|\beta|^2)}}{(\pi|\beta|^2)^{1/4}}\right)\left(\frac{e^{-(q'-|\beta|^2/2)^2/(2|\beta|^2)}}{(\pi|\beta|^2)^{1/4}}\right)
\end{align}
where $e^{i\theta}$ is the phase of $\beta$. Here we recognize the form of a quadrature eigenstate and we combine the two normal distributions to get the much simpler form,
\begin{equation}
\begin{aligned}
      M_{p',q'}  = &\<x_\theta|\frac{(-1)^qe^{i\theta(p'+q')}}{|\beta|(\pi)^{1/4}}
      e^{-(p'+q'-|\beta|^2)^2/(4|\beta|^2)}
\end{aligned}
\end{equation}
Now we see that $p'$ and $q'$ only appear in terms of the sum and difference variable so we can completely move to those variables, including the Jacobian terms we get
\begin{equation}
M_{x,w}\sqrt{dxdw} = \<x_\theta|e^{iw\theta}(-1)^{q'}\frac{e^{-(w-|\beta|^2)^2/(4|\beta|^2)}}{(2\pi)^{1/4}\sqrt{|\beta|}}\sqrt{dxdw}
\end{equation}
Now we move to the POVMs where the phase terms will cancel yielding a very simple form
\begin{equation}
     dxdwE_{x,w} = \frac{e^{-(w-|\beta|^2)^2/(2|\beta|^2)}}{\sqrt{(2\pi)|\beta|^2}} |x_\theta\>\<x_\theta|dxdw
\end{equation}
where we can see that after marginalizing over $w$ we would get 
\begin{equation}
    E_x = |x_\theta\>\<x_\theta|
\end{equation}
as expected.

% ========================================================
\section{Photon number considerations}\label{app:largeLO}
% ========================================================

When discussing the large-LO limit we can break the signal photons into the two modes and compare the number of signal photons that fall into the LO mode to the total number of photons from the LO. There is a subtlety here because we effectively saying that $\<\psi_\sig|\hat{n}(\xi_\sig)|\psi_\sig\> =\<\psi_{\xi_\lo}|\hat{n}(\xi_\lo)|\psi_{\xi_\lo}\> +\<\psi_{\xi_\perp}|\hat{n}(\xi_\perp)|\psi_{\xi_\perp}\> $, which is surprisingly nontrivial, because as an operator equation $\hat{n}(\xi_\sig)\ne\hat{n}(\xi_\lo) +\hat{n}(\xi_\perp) $. We know that the first equation must be true because it says that the total number of photons in the signal is equal to the number of photons from the signal in the LO mode plus the number of photons from the signal in the perpendicular mode. Since the signal mode is described by a linear superposition of just those two modes we know that no photons could fall in any other mode in our decomposition. 

We can also provide evidence that it is true by taking an example of a coherent state signal. We use the decomposition of \cref{eqn:cohdecomp} before taking the expectation value,
\begin{align}
       \<\alpha_{\xi_\sig}|\hat{n}(\xi_s)|\alpha_{\xi_\sig}\> =& \<\gamma\alpha_{\xi_\lo}|\hat{n}(\xi_\lo)|\gamma\alpha_{\xi_\lo}\> + \nonumber\\
       &\<\sqrt{1-|\gamma|^2}\alpha_{\xi_\perp}|\hat{n}(\xi_\perp)|\sqrt{1-|\gamma|^2}\alpha_{\xi_\perp}\>
\end{align}
and 
\begin{equation}
|\alpha|^2 = |\gamma|^2|\alpha|^2 + (1-|\gamma|^2)|\alpha|^2.
\end{equation}
A proof of this fact for any signal state is more subtle, but it begins by introducing an auxiliary mode with no photons in it so that we can treat the change of mode basis thoroughly. This is always allowed since we are extending our single mode basis to include other modes which have no weight. 

We write the signal as $|\pi_\sig\> = |\psi_{\xi_\sig}\>\otimes|0(\xi_{\text{aux}})\>$. We can now decompose our mode operators in the normal way where $\hat{A}(\xi_\sig) \to \gamma\hat{A}(\xi_\lo)+\sqrt{1-|\gamma|^2}\hat{A}(\xi_\sig)$, only now we add in the auxiliary mode so we can write the change of basis as the action of a unitary,
\begin{equation}
    \begin{bmatrix}
        A(\xi_\sig)\\
        A(\xi_{\text{aux}})
    \end{bmatrix}
    =\begin{bmatrix}
        \gamma & \sqrt{1-|\gamma|^2}\\
        \sqrt{1-|\gamma|^2} & -\gamma^* 
    \end{bmatrix}
    \begin{bmatrix}
        A(\xi_\lo)\\
        A(\xi_\perp)
    \end{bmatrix}
\end{equation}
where we have filled in the bottom row by requiring the matrix to be unitary. Now we can calculate $\hat{A}^\dagger(\xi_\sig)\hat{A}(\xi_\sig)$ and see that there are terms present that depend on both modes,
\begin{widetext}
\begin{equation}
    \hat{A}^\dagger(\xi_\sig)\hat{A}(\xi_\sig) = |\gamma|^2\hat{A}^\dagger(\xi_\lo) \hat{A}(\xi_\lo) + (1-|\gamma|^2)\hat{A}^\dagger(\xi_\perp)\hat{A}(\xi_\perp) + \gamma\sqrt{1-|\gamma|^2}\hat{A}^\dagger(\xi_\lo)\hat{A}(\xi_\perp)+\gamma^*\sqrt{1-|\gamma|^2}\hat{A}^\dagger(\xi_\perp)\hat{A}(\xi_\lo). \nonumber
\end{equation}
Similarly we carry out the same calculation for the auxiliary mode
\begin{equation}
    \hat{A}^\dagger(\xi_{\text{aux}})r \hat{A}(\xi_{\text{aux}}) = |\gamma|^2\hat{A}^\dagger(\xi_\perp) \hat{A}(\xi_\perp) + (1-|\gamma|^2)\hat{A}^\dagger(\xi_\lo)\hat{A}(\xi_\lo) - \gamma\sqrt{1-|\gamma|^2}\hat{A}^\dagger(\xi_\lo)\hat{A}(\xi_\perp)-\gamma^*\sqrt{1-|\gamma|^2}\hat{A}^\dagger(\xi_\perp)\hat{A}(\xi_\lo). \nonumber
\end{equation}
We now take expectations of the auxiliary mode $\< \hat{A}^\dagger(\xi_{\text{aux}}) \hat{A}(\xi_{\text{aux}})\> = 0$ and we get the following condition,
\begin{equation}
    \<|\gamma|^2\hat{A}^\dagger(\xi_\perp) \hat{A}(\xi_\perp) + (1-|\gamma|^2)\hat{A}^\dagger(\xi_\lo)\hat{A}(\xi_\lo) \>= \<\gamma\sqrt{1-|\gamma|^2}\hat{A}^\dagger(\xi_\lo)\hat{A}(\xi_\perp)+\gamma^*\sqrt{1-|\gamma|^2}\hat{A}^\dagger(\xi_\perp)\hat{A}(\xi_\lo)\>.
\end{equation}
Now take expectation in the signal mode,
\begin{equation}
     \<\hat{A}^\dagger(\xi_\sig)\hat{A}(\xi_\sig) \>= \<|\gamma|^2\hat{A}^\dagger(\xi_\lo) \hat{A}(\xi_\lo) + (1-|\gamma|^2)\hat{A}^\dagger(\xi_\perp)\hat{A}(\xi_\perp)\> + \<\gamma\sqrt{1-|\gamma|^2}\hat{A}^\dagger(\xi_\lo)\hat{A}(\xi_\perp)+\gamma^*\sqrt{1-|\gamma|^2}\hat{A}^\dagger(\xi_\perp)\hat{A}(\xi_\lo)\>.
\end{equation}
\end{widetext}
Finally just plug in the condition we derived above to get 
\begin{equation}
    \<\hat{A}^\dagger(\xi_\sig)\hat{A}(\xi_\sig) \> = \<\hat{A}^\dagger(\xi_\perp)\hat{A}(\xi_\perp)\> + \<\hat{A}^\dagger(\xi_\lo)\hat{A}(\xi_\lo)\>,
\end{equation}
as desired.

% ========================================================
\section{Combination Rule for POVM}\label{app:convolution}
% ========================================================
Starting from the total Kraus operator in $n$ and $m$ variables \cref{eqn:convolution},
\begin{equation}\label{eqn:combination}
     M_{n,m}= \sum_{p,q} M_{p,q}(\xi_\lo)\otimes M_{n-p,m-q}(\xi_{\perp})\, .
\end{equation}
We can rewrite the order of this sum in terms of discrete sum and difference variables,
\begin{equation}
     M_{n,m}= \sum_{p+q= 0}^{n+m}\,\sum_{p-q = x_{\rm min}}^{x_{\rm max}}M_{p,q}(\xi_\lo)\otimes M_{n-p,m-q}(\xi_{\perp}),
\end{equation}
where $x_{\rm min} = \text{max}(-w,-n,-m)$ and $x_{\rm max} = -x_{\rm min}$. This sum can be approximated quite well in the limit where the difference variable is much less than the sum variable, i.e. $x\ll w$, which is almost always the case for quadrature detection,
\begin{equation}
     M_{n,m}= \sum_{p+q= 0}^{n+m}\,\,\sum_{p-q = -p+q}^{p+q}M_{p,q}(\xi_\lo)\otimes M_{n-p,m-q}(\xi_{\perp}) \, .
\end{equation}
Now move to the sum and difference variables, $x = (p-q)/\sqrt{2}|\beta|$, $w = p+q$. Since we have scaled the difference variable so that it is small, we can approximate the sum over it as an integral. Applying all this in the large LO limit gives
\begin{equation}
    M_{x,w} = \sum_{w'} \int_{-\infty}^{\infty} \frac{|\beta|dx'}{\sqrt{2}} M_{x',w'}(\xi_\lo)\otimes M_{x -x',w-w'}(\xi_{\perp})\, .
\end{equation}
This demonstrates that the total Kraus operator is a convolution of the two constituent Kraus operators. We now need to determine how this convolution changes when we move to the POVMs. This can be done by direct computation but the necessary orthogonality relations are more clear before we move to the sum and difference variables so we will start again from \cref{eqn:convolution}
\begin{widetext}
\begin{equation}
    \begin{aligned}
        E_{n,m} = M_{n,m}^\dagger M_{n,m}
        &= \sum_{p,q} M^\dagger_{p,q}(\xi_\lo) M^\dagger_{n-p,m-q}(\xi_{\perp})\sum_{p',q'} M^\dagger_{p',q'}(\xi_\lo) M^\dagger_{n-p',m-q'}(\xi_{\perp})\, .\\
        &= \sum_{p,q}\sum_{p',q'} M^\dagger_{p,q}(\xi_\lo) (\xi_{\perp}) M_{p',q'}(\xi_\lo) M^\dagger_{n-p,m-q}M_{n-p',m-q'}(\xi_{\perp})\, .
    \end{aligned}
\end{equation}
Now we need to remember the form of the single mode Kraus operators using \cref{eqn:POVMseperated137},  
\begin{equation}
    \begin{aligned}
        M^\dagger_{p,q}(\xi_{\lo}) (\xi_{\perp}) M_{p',q'}(\xi_{\lo})  &= \<\beta|U_{\text{BS}}^\dagger|p\>|q\>\<p|\<q||q'\>|p'\>\<q'|\<p'|U_{\text{BS}}|\beta\>
         = \<\beta|U_{\text{BS}}^\dagger|p\>|q\>\<q|\<p|U_{\text{BS}}|\beta\>\delta_{p,p'}\delta_{q,q'},
    \end{aligned}
\end{equation}
\end{widetext}
where we have applied the orthogonality of Fock states. A similar identity holds for the perpendicular mode. Applying this to the POVM reduces the four sums to a sum over just two variables so we get 
\begin{equation}
\begin{aligned}
    E_{n,m} &= \sum_{p,q} M^\dagger_{p,q}(\xi_{\lo})M_{p,q}(\xi_{\lo}) M^\dagger_{n-p,m-q}M_{n-p,m-q}(\xi_{\perp})\\
    &=\sum_{p,q} E_{p,q}(\xi_\lo)E_{n-p,m-q}(\xi_\perp)\, .
\end{aligned}
\end{equation}
From here we notice that this matches the form of what we started with in \cref{eqn:combination} so we can assert the combination rule for the POVMs in terms of the sum and difference variable is 
\begin{equation}
    E_{x,w} = \sum_{w'} \int_{-\infty}^{\infty} \frac{|\beta|dx'}{\sqrt{2}} E_{x',w'}(\xi_{\lo})\otimes E_{x -x',w-w'}(\xi_{\perp})\, .
\end{equation}

% ========================================================
\section{Time Dependent Photo Record}\label{app:time_dep}
% ========================================================
If we now assume that our detector does produce timing information then we can still get the same answer but our measurement operators must change. We still want to average the time-dependent photo-record which will be accomplished by coarse-graining over time. First we must make some assumptions about our detector. We will model the detector time dependence by saying each detection even is contained in a time bin $(t_i,t_i+\Delta t)$. We will also assume that $\Delta t $ is small with respect to the total detection time $T$.  

With these assumptions, we can write a corrected form of our detector.
\begin{equation}\label{eqn:Time_dep_detector}
\begin{aligned}
     |n\>_{D}(t_i) &= |0\>\otimes\dots |0\>\otimes|n\>\otimes|0\>\otimes \dots |0\>\\
                   &\equiv |n_i\>
\end{aligned}
\end{equation}
This represents getting $n$ clicks in the $i^{\text{th}}$ time bucket. With these, we can write the most general measurement operator for our apparatus as 
\begin{equation}\label{eqn:time_dependent_kraus}
    M_{n,m}(t_i,t_j) = \<n_i|\<m_j|U_{\text{BS}}.
\end{equation}
which corresponds to $n$ clicks on one detector in the $i^{\text{th}}$ and $m$ clicks in the other in the $j^{\text{th}}$ bucket. Up to this point, our measurement operators are fully time-dependent and indeed this analysis could be continued without coarse-graining, but the resulting theory is difficult to parse analytically and seems more suited for numerics. For this reason, we will consider the case of the averaged photo current which requires coarse-graining over time.

Once we have the measurement operators we can assemble the POVMs. The fully time-dependent POVMs would be 
\begin{equation}
    E_{n,m}(t_i,t_j) = M_{n,m}^\dagger(t_i,t_j)M_{n,m}(t_i,t_j)
\end{equation}
and the naively coarse-grained POVMs should be
\begin{equation}
\begin{aligned}
    E_{n,m} &= \sum_{i,j} M_{n,m}^\dagger(t_i,t_j)M_{n,m}(t_i,t_j)\\
            &= \sum_{i,j} U_{\text{BS}}^\dagger|n_i\>|m_j\>\<n_i|\<m_j|U_{\text{BS}}
\end{aligned}
\end{equation}

but this only accounts for the cases were all $n$ and $m$ clicks were in a single time bin. It should also be possible to get say $n/2$ clicks in the first bin and $n/2$ in the second bin for a total of $n$ clicks. So we need to add in these terms.

\begin{equation}
\begin{aligned}
    E_{n,m} &= \sum_{n_i}^n\sum_{m_j}^m U_{\text{BS}}^\dagger|n_i\>|m_j\>\<n_i|\<m_j|U_{\text{BS}}\\ 
\end{aligned}
\end{equation}
where $\sum_{n_i}^n$ indicates a sum over all possible values of $n_1,n_2,\dots$ such that $\sum_i n_i = n$. An illustrative special case is when there are only two detection windows, $(t_1,t_2)$ and we get:
\begin{widetext}
\begin{equation}
\begin{aligned}
    E_{n,m} &= \sum_{p}\sum_{q} U_{\text{BS}}^\dagger\left(|p\>_1\otimes|n-p\>_2\right)\left(|q\>_1\otimes|m-q\>_2\right)\left(\<p|_1\otimes\<n-p|_2\right)\left(\<q|_1\otimes\<m-q|_2\right)U_{\text{BS}}\\ 
\end{aligned}
\end{equation}
Now we note that POVMs are basis independent i.e the measurement statistics are the same regardless of any change of bases we make on the POVMs.  So we can conjugate our POVM by some unitary $U$ so that we move to the Gram-Schmidt basis defined in \cref{eqn:basis}. This is where we will assume that $\Delta t\ll T$ so that our bin basis spans the same set of functions as our Gram-Schmidt basis. Now we have 
\begin{align}
   V^\dagger E_{n,m}V = \sum_{p}\sum_{q} U_{\text{BS}}^\dagger & \left(|p_{\xi_\lo}\>\otimes|n-p_{\xi_\perp}\>\right)\left(|q_{\xi_\lo}\>\otimes|m-q_{\xi_\perp}\>\right)  
    \left(\<p_{\xi_\lo}|\otimes\<n-p_{\xi_\perp}|\right)\left(\<q_{\xi_\lo}|\otimes\<m-q_{\xi_\perp}|\right)U_{\text{BS}}\, ,
\end{align}
\end{widetext}
where $V$ denotes a change of basis from the numbered modes to the Gram-Schmidt basis. The final expression is equivalent to what we had before. Note that here we are able to limit ourselves to the special case of just two modes because we are omitting the trivial modes that complete our basis, but carry 0 photons.

% ========================================================
\section{Coherent State Difference Variable Distribution Calculations}\label{app:x_var}
% ========================================================
Using the total POVM from \cref{eqn:POVMseperated137} we can derive the distribution of the total measurement by taking the expectation of the POVM in the signal state,
\begin{align}
P(x) &= \<\alpha_{\xi_\sig}|E_x|\alpha _{\xi_\sig}\> \nonumber\\  
 &=  \int dx'  \<\gamma\alpha|E_{x'}(\xi_\lo)|\gamma\alpha\> \times \nonumber\\
&\quad  \quad\quad  \<\sqrt{1-|\gamma|^2}\alpha| E_{x-x'}(\xi_{\perp})|\sqrt{1-|\gamma|^2}\alpha\> \, .\nonumber 
\end{align}
This distribution can naturally be decomposed into two parts, and then combined by a convolution. 

First, let's consider the component in the local oscillator mode. The probability distribution is given by $P(x) = \<\gamma\alpha_{\xi_\lo}|E^\beta_{x,\xi_\lo}|\gamma\alpha_{\xi_\lo}\>$ i.e.
\begin{equation}
\begin{aligned}
    P(x) = |\<x|\gamma\alpha\>|^2
    &= \frac{e^{-x^2}}{\sqrt{\pi}}|\<0|e^{\sqrt{2}x\hat{a}}e^{-\hat{a}^2/2}|\gamma\alpha\>|^2\\
    &= \frac{e^{-x^2}}{\sqrt{\pi}}|e^{-|\gamma\alpha|^2/2}e^{\sqrt{2}\gamma\alpha}e^{-\gamma^2\alpha^2/2}|^2\\
    &= e^{-|\gamma\alpha|^2}\frac{e^{-x^2}}{\sqrt{\pi}}e^{2\sqrt{2}\text{Re}(\gamma\alpha)}e^{-\text{Re}(\gamma^2\alpha^2)} \, .
\end{aligned}
\end{equation}
Here we should recall our convention that $\alpha$ is real and that the phase is completely contained in the modes and thus in $\gamma$. We can introduce shorthand $\gamma = \gamma_R + i\gamma_I$. 

Now we can complete the square to obtain a Gaussian distribution in $x$ that is not mean 0
\begin{equation}
    -(x^2-2\sqrt{2}x\alpha\gamma_R) = -(x-\sqrt{2}\alpha\gamma_R)^2 + 2\alpha^2\gamma_R^2.
\end{equation}
With this we can rewrite the distribution 
\begin{equation}
    P(x) = \frac{e^{-(x-\sqrt{2}\alpha\gamma_R)^2}}{\sqrt{\pi}}\exp\big[-\alpha^2(|\gamma|^2+\gamma_R^2- \gamma_I^2-2\gamma_R^2)\big]
\end{equation}
where we have used the fact that $\text{Re}(\gamma^2) = \gamma_R^2- \gamma_I^2$. It can be shown in a couple lines of algebra that $|\gamma|^2+\gamma_R^2- \gamma_I^2-2\gamma_R^2=0$ and so we end up with just a normal distribution in $x$.
\begin{equation}
    P(x) = \frac{e^{-(x-\sqrt{2}\alpha\gamma_R)^2}}{\sqrt{\pi}}
\end{equation}

For the perpendicular mode we start with \cref{eqn:seperated}
\begin{widetext}
\begin{equation}
     P(x,w)_\perp = \<\sqrt{1-|\gamma|^2}\alpha|\sum_{w} |w_{\xi_\perp}\>\<w_{\xi_\perp}|\text{Bin}\Big(\frac{|\beta|x}{\sqrt{2}}\Big|w,\frac{1}{2},0\Big)|\sqrt{1-|\gamma|^2}\alpha\>\\
\end{equation}
where we can apply the explicit formula for the shifted binomial distribution. At this point both $x$ and $w$ are discrete so we will need to move to continuous variables after we simplify the expression. After taking expectation in the coherent state we have,
\begin{equation}
    P(x,w)_\perp = \frac{w!}{\left(\frac{|\beta|x}{\sqrt{2}}+\frac{w}{2}\right)!\left(\frac{w}{2}-\frac{|\beta|x}{\sqrt{2}}\right)!}\frac{1}{2^w}\frac{((1-|\gamma|^2)|\alpha|^2)^we^{-(1-|\gamma|^2)|\alpha|^2}}{w!}
\end{equation}
which is the product of the shifted binomial and the Poisson distribution. We can simplify this into the product of two Poisson distributions as 
\begin{equation}
    P(x,w)_\perp = \left(\frac{(1-|\gamma|^2)|\alpha|^2}{2}\right)^{w/2 - |\beta|x/\sqrt{2}}\frac{e^{-(1-|\gamma|^2)|\alpha|^2/2}}{\left(\frac{w}{2}-\frac{|\beta|x}{\sqrt{2}}\right)!}\left(\frac{(1-|\gamma|^2)|\alpha|^2}{2}\right)^{w/2 + |\beta|x/\sqrt{2}}\frac{e^{-(1-|\gamma|^2)|\alpha|^2/2}}{\left(\frac{w}{2}+\frac{|\beta|x}{\sqrt{2}}\right)!}\, .
\end{equation}

No we move to continuous variables by approximating both Poisson distributions as normal, this is valid so long as $(1-|\gamma|^2)|\alpha|^2$ is large enough for the central limit theorem to apply. Doing this an applying some algebraic simplification yields,
\begin{equation}
   P(x,w)_\perp dxdw= \frac{|\beta|dxdw}{\sqrt{2}\pi(1-|\gamma|^2)|\alpha|^2}\exp\left[-\frac{|\beta|^2x^2}{(1-|\gamma|^2)|\alpha|^2}\right]\exp\left[-\frac{(w-(1-|\gamma|^2)|\alpha|^2)^2}{2(1-|\gamma|^2)|\alpha|^2}\right]\, ,
\end{equation}
\end{widetext}
which we can marginalize over $w$ to get the difference variable distribution
\begin{equation}
\begin{aligned}
        P(x)_\perp &= \frac{|\beta|}{\sqrt{\pi(1-|\gamma|^2)|\alpha|^2}}\exp\left[-\frac{|\beta|^2x^2}{(1-|\gamma|^2)|\alpha|^2}\right]\\
        &= \mathcal{N}\left(x,\mu= 0,\sigma^2 = \frac{|\alpha|^2(1-|\gamma|^2)}{2|\beta|^2}\right)\,.
\end{aligned}
\end{equation}

% ========================================================
 \section{Filtering Theory}\label{app:filter}
% ========================================================
In single mode homodyne we know that the measurement reduces to a measurement of a quadrature defined by the phase of the LO. As long as the LO is large enough for the LO shot noise to dominate the signal shot noise we can reduce the effective quadrature noise to that of vacuum fluctuations. For the multimode case we show in \cref{eqn:POVMseperated}, that the measurement is a quadrature measurement convolved with additional intensity noise from the measurement of the mismatched portion of the signal. The goal of filtering in this context is to reduce the intensity like noise present in our time-averaged outcome, while not affecting the quadrature measurement at all.

We will first imagine that we have very fast detectors that collect photons in a very small time bin $d\tau$ for each data point. Over the entire detection interval, $T$ we will assume we have many data points i.e $d \tau \ll T$. We will consider the filtering operation as applying some time-dependent set of weights $f(t)$ to the photocurrent, and then averaging over the filtered data. This means that any outcome of our measurement $x$ is given by 
\begin{equation}
    x = \int_0^T f(t)x(t)dt,
\end{equation}
where we can easily replace this integral with a sum if the detector window $d\tau$ is not infinitesimal.

In order to ensure that the operator being measured is $\hat{Q}(\xi_\lo)$ any filter we apply must be constant over the LO mode. If this is not the case then the measurement will have reduced sensitivity to the portions of the LO mode when $f(t)$ is small. The extreme example of this is when $f(t)=0$ on some interval $(t_0,t_1)$, clearly since these data points are completely removed from the final measurement outcome our measurement has no sensitivity to the part of the LO mode. So we conclude that the filter must leave the LO mode unchanged so the measurement is maintained , i.e $f(t)\xi_\lo(t) \propto \xi_\lo(t)$

The set of all possible filters $f(t)$ under these restrictions becomes
\begin{equation}
    f(t) = \begin{cases} 
      c & \text{if } \xi_{\lo}(t) \ne 0  \\
      g(t) & \text{if } \xi_\lo(t) = 0
   \end{cases}.
\end{equation}
where $c$ is some constant. Without loss of generality we will assume that $c=1$ because any other choice of constant would merely scale the value of all outcomes, leaving the SNR unchanged. The problem is now to find $g(t)$ so that we have the minimum perpendicular mode noise in our measurement. 

At this point we should note that in many cases $\xi_\lo(t)$ is never zero. While ultimately this means that there is no filter that will leave the LO mode unchanged, it may still be desirable to find an approximate filter which greatly increases SNR at the cost of a small change in the measurement. For example if the LO is a Gaussian pulse then 5 standard deviations away from the mean might be sufficient to approximate $\xi_\lo(t) \approx 0 $. If we want to ensure that some small fraction $p$ of the total LO mode photons are excluded by our gate then we have the condition
\begin{equation}
    |\xi_\lo(t)|^2< \frac{p}{d\tau}.
\end{equation}

Lets now consider the case where the signal is a coherent state $|\alpha_{\xi_\sig}\>$. Here semiclassical analysis tells us how the mean and variance will change under the proposed filter,
\begin{equation}
    \begin{aligned}
        \mu &\to 2\text{Re}(\alpha\beta^*)\text{Re}(\int_0^T f(t)\xi_\lo^*(t)\xi_s(t)dt)\\
        \sigma^2 &\to |\beta|^2\int_0^T|f(t)|^2|\xi_\lo(t)|^2dt + |\alpha|^2\int_0^T|f(t)|^2|\xi_\sig(t)|^2dt.
    \end{aligned}
\end{equation}
Now we apply the fact that $f(t)\xi_\lo(t) = \xi_\lo(t)$ and the convention that $\beta$ is real to get the simplified mean 
\begin{equation}
    \mu = 2|\beta|\text{Re}(\gamma\alpha),
\end{equation}
and  the variance
\begin{align}
\sigma^2 &= |\beta|^2 + |\alpha|^2\int_0^T|f(t)|^2|\xi_\sig(t)|^2 \nonumber\\
       &= |\beta|^2 + |\alpha|^2 \int_0^T|f(t)|^2|\gamma\xi_\lo + \sqrt{1-|\gamma|^2}\xi_\perp|^2.
\end{align}
We can expand the second term (the one proportional to $|\alpha|^2$ to $\int_0^T|f(t)|^2 [|\gamma\xi_\lo|^2 + (1-|\gamma|^2)|\xi_\perp|^2+ \gamma\sqrt{1-|\gamma|^2}(\xi_\lo^*\xi_\perp + \xi_\lo\xi_\perp^*) ]$ which eventually gives
\begin{align}
\sigma^2 & = |\beta|^2 + |\alpha|^2\left[|\gamma|^2+ (1-|\gamma|^2)\int_0^T|f(t)|^2|\xi_\perp|^2 \right] \, ,
\end{align}
where we have again used that fact that $f\xi_\lo = \xi_\lo$ and that $\<\xi_\lo,\xi_\perp\> = 0 $. Finally we can apply the assumption that $|\beta|^2 \gg |\gamma\alpha|^2$ to get
\begin{equation}
    \text{SNR} = \frac{\mu^2}{\sigma^2} = \frac{4|\beta|^2\text{Re}(\gamma\alpha)}{|\beta|^2 + \eta_f|\alpha|^2}
\end{equation}
where $\eta_f = \int_0^T|f(t)|^2|\xi_\perp|^2$

\section{Single Photon Distribution}\label{app:1photon}
We start the calculation from the input state in \cref{eqn:singlephoton},
\begin{equation}
    |\psi\>_s = \gamma |1_{\xi_\lo}\>|0_{\xi_\perp} \>+\sqrt{1-|\gamma|^2}|0_{\xi_\lo}\>|1_{\xi_\perp}\> \,. 
\end{equation}
Taking expectation of the POVM would produce four terms because the input state is written as two terms. 
\begin{widetext}
\begin{equation}
\begin{aligned}
    P(x) &=    dx \int dx' \<\psi_s| E_{x'}^\lo(\xi_\lo) \otimes  E_{x-x'}^\perp(\xi_{\perp})|\psi_s\> \, ,\\
    &= \int dx'\left(\gamma^*\<1|\<0| + \sqrt{1-|\gamma|^2}\<0|\<1|\right) E_{x'}^\lo \otimes  E_{x-x'}^\perp|\left(\gamma|1\>|0\>+\sqrt{1-|\gamma|^2}|0\>|1\>\right),
\end{aligned}
\end{equation}
where we have simplified the mode labels for conciseness. Now we just note that $\<m_{\xi_\perp}|E^\perp|n_{\xi_\perp}\>\propto \delta_{m,n}$ because it is a projection onto Fock states, and so only two terms survive,
\begin{equation}
\begin{aligned}
    P(x) &= \int dx'|\gamma|^2\<1|E_{x'}^\lo|1\>\<0|E_{x-x'}^\perp|0\> + (1-|\gamma|^2)\<0|E_{x'}^\lo|0\>\<1|   E_{x-x'}^\perp|1\>\\
    &= \int dx'\frac{|\gamma|^2}{2\sqrt{\pi}}e^{-x'^2}H_1^2(x')\text{Bin}\Big(\frac{|\beta|(x-x')}{\sqrt{2}}\Big|0,\frac{1}{2},0\Big) + \frac{(1-|\gamma|^2)}{\sqrt{\pi}}e^{-x'^2}\text{Bin}\Big(\frac{|\beta|(x-x')}{\sqrt{2}}\Big|1,\frac{1}{2},0\Big)
\end{aligned}
\end{equation}
Now we are in a regime of very weak signal where we can not approximate these binomial distributions as normal, instead we will mimic the discrete distribution by restricting the distribution of  $(x-x')$ to only discrete values. In this case that can best be done by approximating the binomial as a sum of delta functions,
\begin{equation}
\begin{aligned}
    P(x) &= \int dx'\frac{|\gamma|^2}{2\sqrt{\pi}}e^{-x'^2}H_1^2(x')\delta(x-x') + \frac{(1-|\gamma|^2)}{\sqrt{\pi}}e^{-x'^2}\left(\frac{1}{2}\delta\left(x-x'-\frac{1}{\sqrt{2}|\beta|}\right) + \frac{1}{2}\delta\left(x-x'+\frac{1}{\sqrt{2}|\beta|}\right)\right)\\
    &= \frac{|\gamma|^2}{2\sqrt{\pi}}e^{-x^2}H_1^2(x)  +\frac{(1-|\gamma|^2)}{2\sqrt{\pi}}\left(e^{-(x-1/\sqrt{2}|\beta|)^2} + e^{-(x+1/\sqrt{2}|\beta|)^2} \right)\\
    &=\frac{1}{2\sqrt{\pi}}\left[4|\gamma|^2e^{-x^2}x^2  +(1-|\gamma|^2)\left(e^{-(x-1/\sqrt{2}|\beta|)^2} + e^{-(x+1/\sqrt{2}|\beta|)^2} \right)\right].
\end{aligned}
\end{equation}
\end{widetext}
This matches with the results in \cref{eqn:single_dist}

% ========================================================
\section{Comparison of Proposed LO Mode Quadrature Projection Limit vs. Filtering Demonstrations}
% ========================================================
Here we compare our conjectured heterodyne quantum limit set by quadrature projection noise in the LO mode with filtering demonstrations \cite{Deschenes2013} and \cite{Deschenes2015}. First we assume a Sech squared temporal intensity profile. Omitting complex phase, since here we are only concerned with the absolute mode overlap:
\begin{equation}
    |\xi_\sig| = \frac{1}{\sqrt{T}},~~~~~~|\xi_\lo| = \frac{1}{\sqrt{2\tau}} \sech \left (\frac{t}{\tau}\right).
\end{equation}
Here $\tau$ is the temporal width of the Sech function. The squared mode overlap $|\big<\xi_\sig, \xi_\lo \big>|^2$ is thus:
\begin{equation}
    |\gamma|^2 = \frac{\pi^2 \tau}{2T}
\end{equation}
The fully general heterodyne SNR--which can be derived from a purely moment-based analysis and makes no assumptions about the relative strength of signal and LO--is:
\begin{equation}
    \frac{\<i_S^2\>}{\<i_{N}^2\>} = \frac{2(\frac{\eta_q e}{h \nu})^2 \frac{P_\lo}{n} P_{s}}{e \frac{\eta_q e}{h \nu} (P_\lo + |\gamma|^2P_\text{s}) B},
\end{equation}
where $\<i_S^2\>$ is the average signal power, $\<i_{N}^2\>$ is the quantum noise solely in the temporal mode of the comb local oscillator, $\eta_q$ is the quantum efficiency of the detector, $e$ is the fundamental charge, $h\nu$ is the energy per photon, $P_\text{LO}$ is the total comb power, $n$ is the number of comb teeth, $P_\text{s}$ is the CW power, and $B$ is the resolution bandwidth. For transform-limited sech pulses $\tau = \frac{0.315} {1.76 \Delta\nu}$.

In \cite{Deschenes2013}, $\nu = 193 ~\text{THz}$, $T = 10 ~\text{ns}$, $\Delta\nu = 32 ~\text{GHz}$, $\eta_q = 0.76$, $P_\text{LO} = 6 ~\text{nW}$, $P_\text{s} = 2.8 ~\text{mW}$, $n = 320$, and $B = 170 ~\text{kHz}$. Our predicted SQL SNR is 54.5 dB versus the experimentally realized SNR of 36.9 dB. In [20], $\nu = 193 ~\text{THz}$, $T = 10 ~\text{ns}$, $\Delta\nu = 12 ~\text{GHz}$, $\eta_q = 0.7$, $P_\text{LO} = 1.74 ~\mu\text{W}$, $P_\text{s} = 0.5 ~\text{mW}$, $n = 120$, and $B = 100 ~\text{kHz}$. Our predicted SQL SNR is 78.6 dB versus the experimentally realized SNR of 68.3 dB. Similar SNR figures arise when assuming a Gaussian profile (54.2 dB and 78.5 dB for \cite{Deschenes2013} and \cite{Deschenes2015}, respectively).

Similarly, we can calculate $\eta_f$ for this experiment using some assumptions. Given the $60$ ps pulse width we can approximate $f(t)$ by moving 5 standard deviations away from the pulse mean. Since the signal was a CW laser the integral to calculate $\eta_f$ is simple and we get the following:
\begin{equation}
    \eta_f \approx \frac{5\sigma}{T}.
\end{equation}
After correctly converting from pulse width to standard deviations of a Gaussian pulse we get $\eta_f = 1.3 \times 10^{-2}$ for \cite{Deschenes2013}. Along with the other experimental details, this would result in a potential $18.8$ dB improvement, slightly less than the 20 dB they achieved. This difference can be explained simply because they did not apply a filter that we consider. Additionally, if we choose to approximate our cutoff after only 3 standard deviations we would have had an improvement of 20.8 dB, which better matches their result.

\end{document}